\documentclass[10pt]{cernrep}
\usepackage{graphicx}
\usepackage{floatflt}
\usepackage{color}

\newcommand{\Pt}{{P_t}}
\newcommand{\dphi}{\Delta\phi}

\newcommand{\Dbgj}{\Pt^{\gamma+Jet}}   

\newcommand{\la}{\langle}
\newcommand{\ra}{\rangle}
\newcommand{\gpj}{~"$\gamma+Jet$"~}
\newcommand{\rrr}{\to} 

\newcommand{\pth}{\hat{p}_{\perp}^{\;min}}
\newcommand{\Db}{$\Pt(O\!\small{+}\!\eta\!>\!\!\!5)$}

\newcommand{\ptg}{$\Pt^{\gamma}$}

\newcommand{\Fptgj}{(\Pt^{\gamma}\!\!-\!\!\Pt^{Jet})/\Pt^{\gamma}}
\newcommand{\Gvc}{\footnotesize{$(GeV/c)$} }

\newcommand{\hn}{\hspace*{-8mm}}
\newcommand{\hm}{\hspace*{-0.6mm}}
\newcommand{\hmm}{\hspace*{-1.3mm}}
\newcommand{\han}{\hspace*{.58cm}}
\newcommand{\has}{\hspace*{.52cm}}
\newcommand{\hass}{\hspace*{.4cm}}
\newcommand{\had}{\hspace*{.54cm}}
\newcommand{\hbn}{\hspace*{.59cm}}
\newcommand{\hbd}{\hspace*{.52cm}}
\newcommand{\hcn}{\hspace*{.57cm}}

\newcommand{\hcd}{\hspace*{.52cm}}

\definecolor{Light}{gray}{0.80}

\suppressfloats[!]

\def\baselinestretch{1.0}

\begin{document}


\thispagestyle{empty}

\vskip-5mm

\begin{center}
{\Large JOINT INSTITUTE FOR NUCLEAR RECEARCH}
\end{center}

\vskip10mm

\begin{flushright}
JINR Preprint \\
E2-2000-255 \\
hep-ex/0011017
\end{flushright}

\vspace*{3cm}

\begin{center}
\noindent
{\Large{\bfseries Jet energy scale setting with \gpj events at LHC \\[0pt]
energies. Detailed study of the background suppression.}}\\[5mm]
{\large D.V.~Bandourin$^{1\,\dag}$, V.F.~Konoplyanikov$^{2\,\ast}$, 
N.B.~Skachkov$^{3\,\dag}$}

\vskip 0mm

{\small
{\it
E-mail: (1) dmv@cv.jinr.ru, (2) kon@cv.jinr.ru, (3) skachkov@cv.jinr.ru}}\\[3mm]
$\dag$ \large \it Laboratory of Nuclear Problems \\
\hspace*{-4mm} $\ast$ \large \it Laboratory of Particle Physics
\end{center}

\vskip 9mm
\begin{center}
\begin{minipage}{150mm}
\centerline{\bf Abstract}
\noindent
The possibilities of the background events suppression, based on
the QCD subprocesses of $qg-,gg-,qq-$ scattering with big cross sections,
to the signal \gpj events are studied. Basing on the introduced selection
criteria, the background suppression factors and signal events selection
efficiencies and the number of the events, that can be collected
at LHC with low luminosity $L=10^{33}cm^2 s^{-1}$ during one month
of continuous work, are determined.  
\end{minipage}
\end{center}

\newpage

\setcounter{page}{1}
\vskip-2mm
\section{INTRODUCTION} 
This paper continues our previous publications [1--4],
 where possibilities of jet energy scale setting and
calibration of the hadron calorimeter at LHC energies by using \gpj process
have been studied.

This article is devoted to the study of the background events suppression and
to demonstration of the efficiency of the cuts introduced by us
in paper [1] for this purpose.

\section{ILLUSTRATION OF NEW CUTS EFFICIENCY}                                            

To estimate the background for the signal events, we have done the simulation
\footnote{
PYTHIA~5.7 version with default CTEQ2L parametrisation of structure functions
is used here.}
with a mixture of all QCD and SM subprocesses existing in PYTHIA with large
cross sections, namely: ISUB=1, 2, 11--20, 28--31, 53, 68, which can lead to
a big background for our main "signal" subprocesses (ISUB=14 and 29 in PYTHIA)
\footnote{A contribution of another possible NLO channel $gg\rrr g\gamma$
(ISUB=115 in PYTHIA) was found to be still negligible even at LHC energies.}:
\begin{eqnarray}
\hspace*{4.9cm} qg\to q+\gamma \hspace*{5.8cm} (1a)
\nonumber
\end{eqnarray}
\vspace{-7mm}
\begin{eqnarray}
\hspace*{4.9cm} q\overline{q}\to g+\gamma  \hspace*{5.8cm} (1b)
\nonumber
\end{eqnarray}
\setcounter{equation}{1}
\vspace{-4mm}

Three generations (each of $50\cdot10^6$
events) with different values of minimal $\Pt$ of hard process
$\pth$
\footnote{CKIN(3) parameter in PYTHIA.}
have been done: the first one is with
$\pth$ = 40 $GeV/c$, while the second and the third -- with $\pth$ = 100 and
 200 $GeV/c$, respectively.

We have selected ``$\gamma$-candidate +1 Jet'' events
 with $\Pt^{jet}> 30~ GeV/c$ containing
one $\gamma$-candidate to be identified by the detector as an
isolated photon with $\Pt^{\gamma}\geq40$ (~100 and ~200) $~GeV/c$
for the generation with $\pth \geq 40$ (~100 and ~200) $~GeV/c$, respectively.
One needs to stress that here and below, speaking about the $\gamma$-candidate,
we imply in reality a signal that may be registered in the
3 by 3 ECAL crystal cell window with the highest $\Pt$ $\gamma/e$ in the center.
All these photon-candidates were supposed to satisfy the isolation criteria
1--4 of Section 3.2 from paper [1] with
$\Pt^{isol}_{CUT} = 2 ~GeV/c$ and $\epsilon^{\gamma}_{CUT}=5\% $.
No special cuts were imposed on
$\Delta\phi$, $\Pt^{out}$ and $\Pt^{clust}$
(the values of $\Pt^{clust}$ are automatically limited from the top since
 we select ``$\gamma$-candidate +1 jet'' events with $\Pt^{Jet}>30~ GeV/c$).

The corresponding distributions for the physical observables,
introduced in Sections 3.1 and 3.2 from [1], are shown separately for
the signal
``$\gamma$ - dir'' and the background events in Figs.~1, 3, 5 and in
scatter plots 2, 4, 6. First columns in these figures, denoted by
''$\gamma$ - dir'', show the distributions in the signal events, i.e. in
the events corresponding to the processes (1a) and (1b), having direct photons
with $\Pt^{\gamma-dir}\geq 40 ~GeV/c$. The second columns, denoted as
''$\gamma$ - brem'', correspond to the events in which the photons
were emitted from quarks (i.e. bremsstrahlung photons)
and  have passed under the imposed cuts. The distributions in the
third column were built basing on the events containing
``$\gamma$-mes'' photons, i.e. those photons that
 originate from multiphoton decays of mesons
($\pi^0$, $\eta$, $\omega$, $K^0_S$) and have also passed under
the imposed cuts.

\begin{figure}[htbp]
 \vspace*{-0.0cm}
 \hspace*{-0.7cm}
  \includegraphics[width=14.cm,height=19cm]{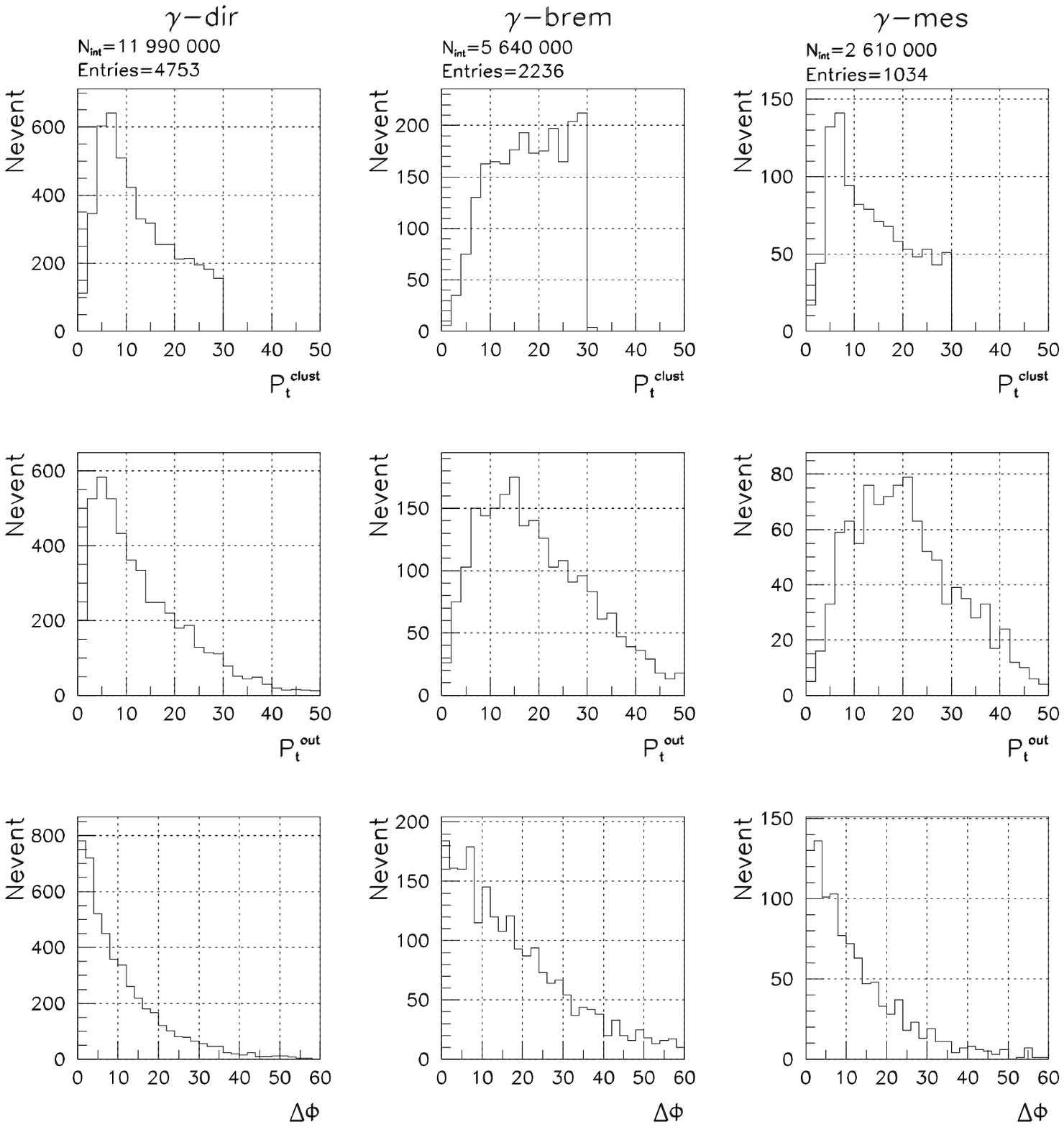}
  \vspace{-0.8cm}
    \caption{\hspace*{0.0cm} \normalsize Signal/Background: Number of events distribution
over $\Pt^{clust}$, $\Pt^{out}$, $\dphi$ ($\Pt^{\gamma}\geq40~GeV/c$)}
    \label{fig:1b40}
  \end{figure}

Firstly we see that in the case of $\Pt^{\gamma} \geq 200 ~GeV/c$
(see Fig.~\ref{fig:1b200}) practically all ``signal events'' have
$\Delta \phi < 15^{\circ}$, and in the case $\Pt^{\gamma} \geq 100 ~GeV/c$
(see Fig.~\ref{fig:1b100}) most of them are also within
$\Delta \phi < 15^{\circ}$.
It is seen from Fig.~\ref{fig:1b40} that at lower values
$\Pt^{\gamma} \geq 40 ~GeV/c$ there is still
a big number (about $70\%$) of the signal events belonging to the interval of
$\Delta \phi < 15^{\circ}$.
From here
\begin{figure}[htbp]
 \vspace{-0.0cm}
 \hspace*{-0.7cm}
  \includegraphics[width=14.cm,height=19cm]{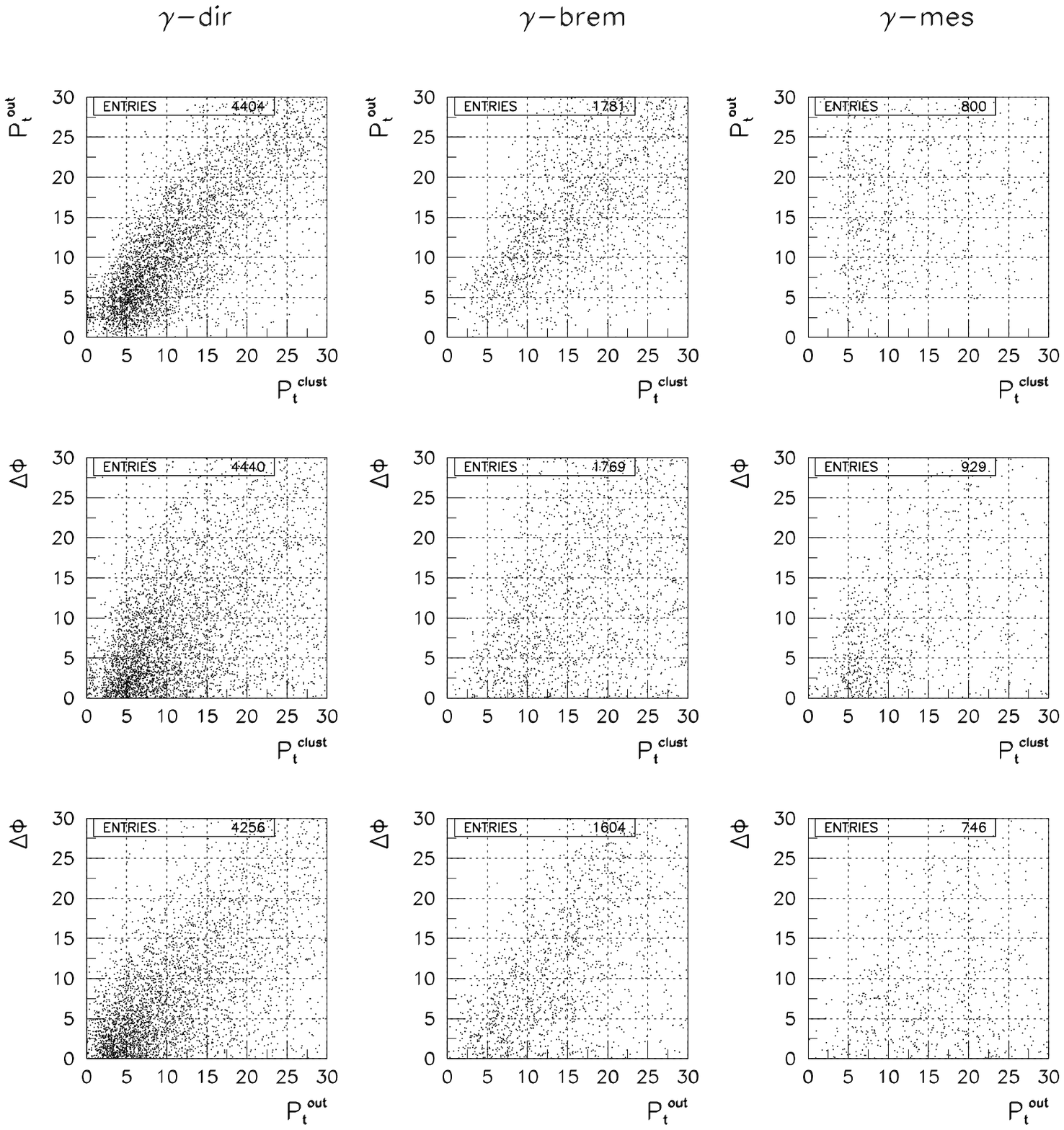}
  \vspace{-0.8cm}
    \caption{\hspace*{0.0cm}\normalsize Signal/Background: $\Pt^{clust}$ vs. $\Pt^{out}$,
$\Pt^{clust}$ vs.$\dphi$, $\Pt^{out}$ vs. $\dphi$ ($\Pt^{\gamma}\geq40~GeV/c$)}
    \label{fig:2b40}
  \end{figure}

\begin{figure}[htbp]
 \vspace{0.0cm}
 \hspace*{-0.7cm}
 \includegraphics[width=14cm,height=19cm]{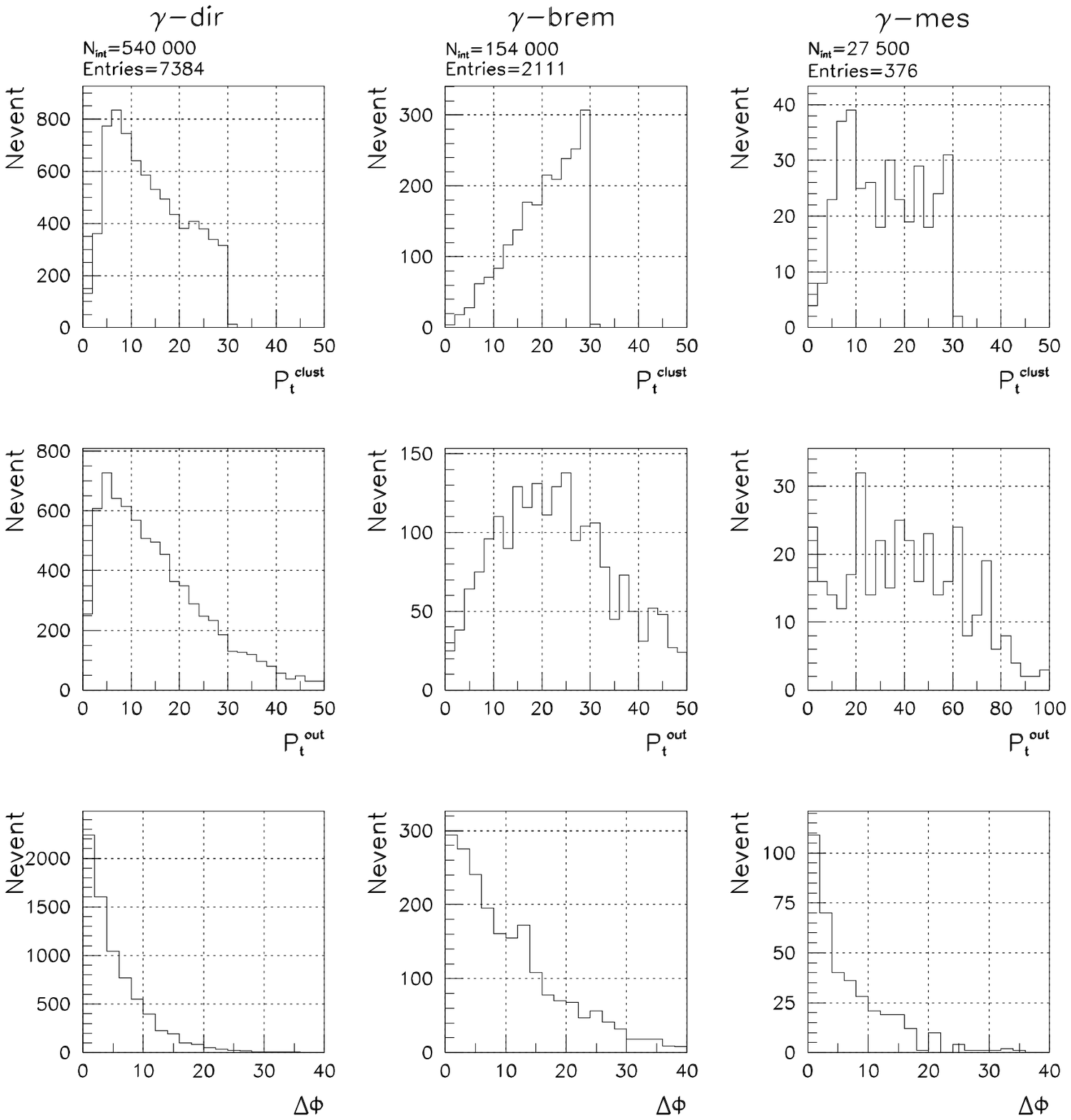}
  \vspace{-0.8cm}
    \caption{\hspace*{0.0cm} \normalsize Signal/Background: Number of events distribution
over $\Pt^{clust}$, $\Pt^{out}$, $\dphi$ ($\Pt^{\gamma}\geq100~GeV/c$)}
    \label{fig:1b100}
  \end{figure}
\begin{figure}[htbp]
 \vspace{0.0cm}
 \hspace*{-0.7cm}
  \includegraphics[width=14cm,height=19cm]{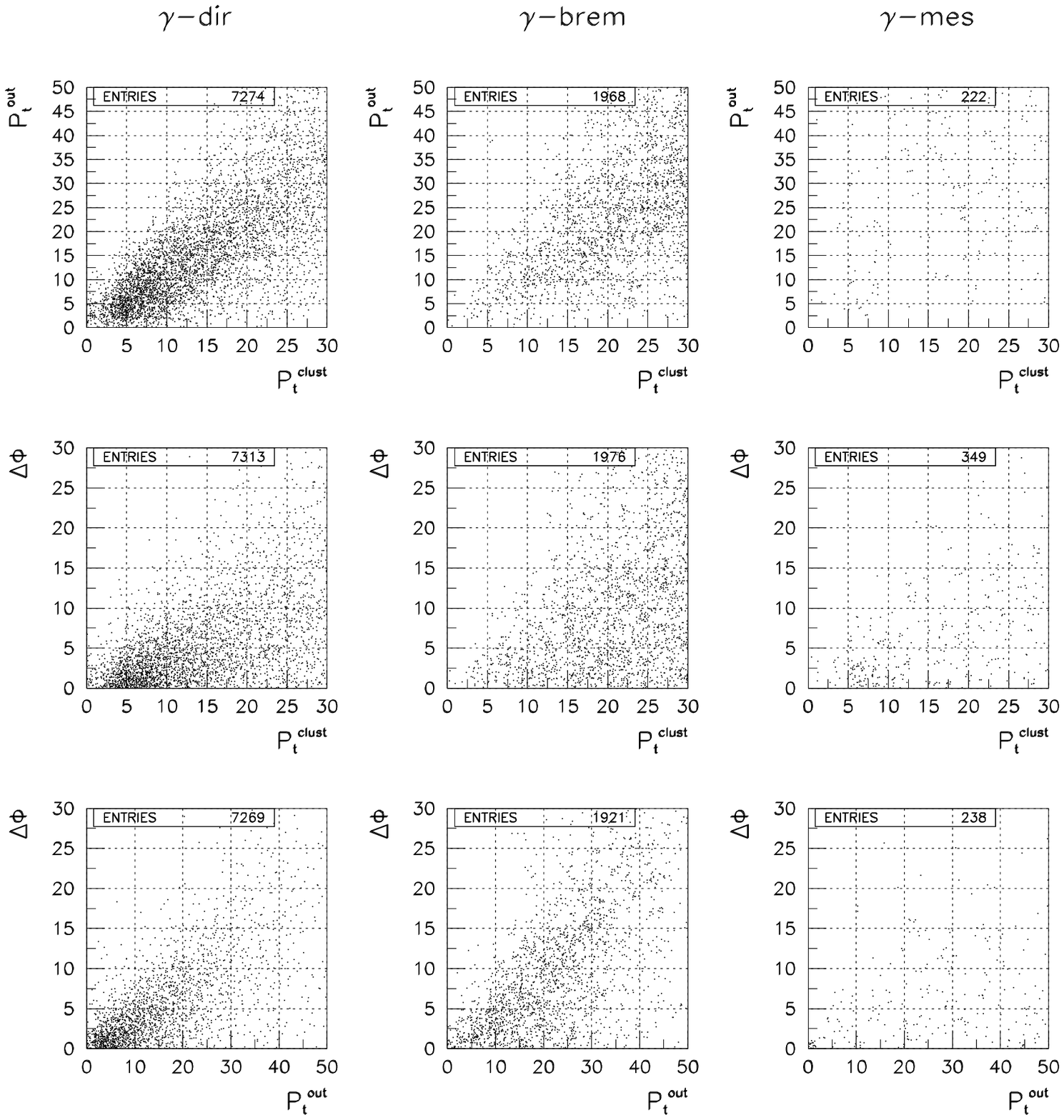}
  \vspace{-0.8cm}
    \caption{\hspace*{0.0cm}\normalsize  Signal/Background: $\Pt^{clust}$ vs. $\Pt^{out}$,
$\Pt^{clust}$ vs.$\dphi$, $\Pt^{out}$ vs. $\dphi$ ($\Pt^{\gamma}\geq100~GeV/c$)}
    \label{fig:2b100}
  \end{figure}

\begin{figure}[htbp]
 \vspace{0.0cm}
 \hspace*{-0.7cm}
 \includegraphics[width=14cm,height=19cm]{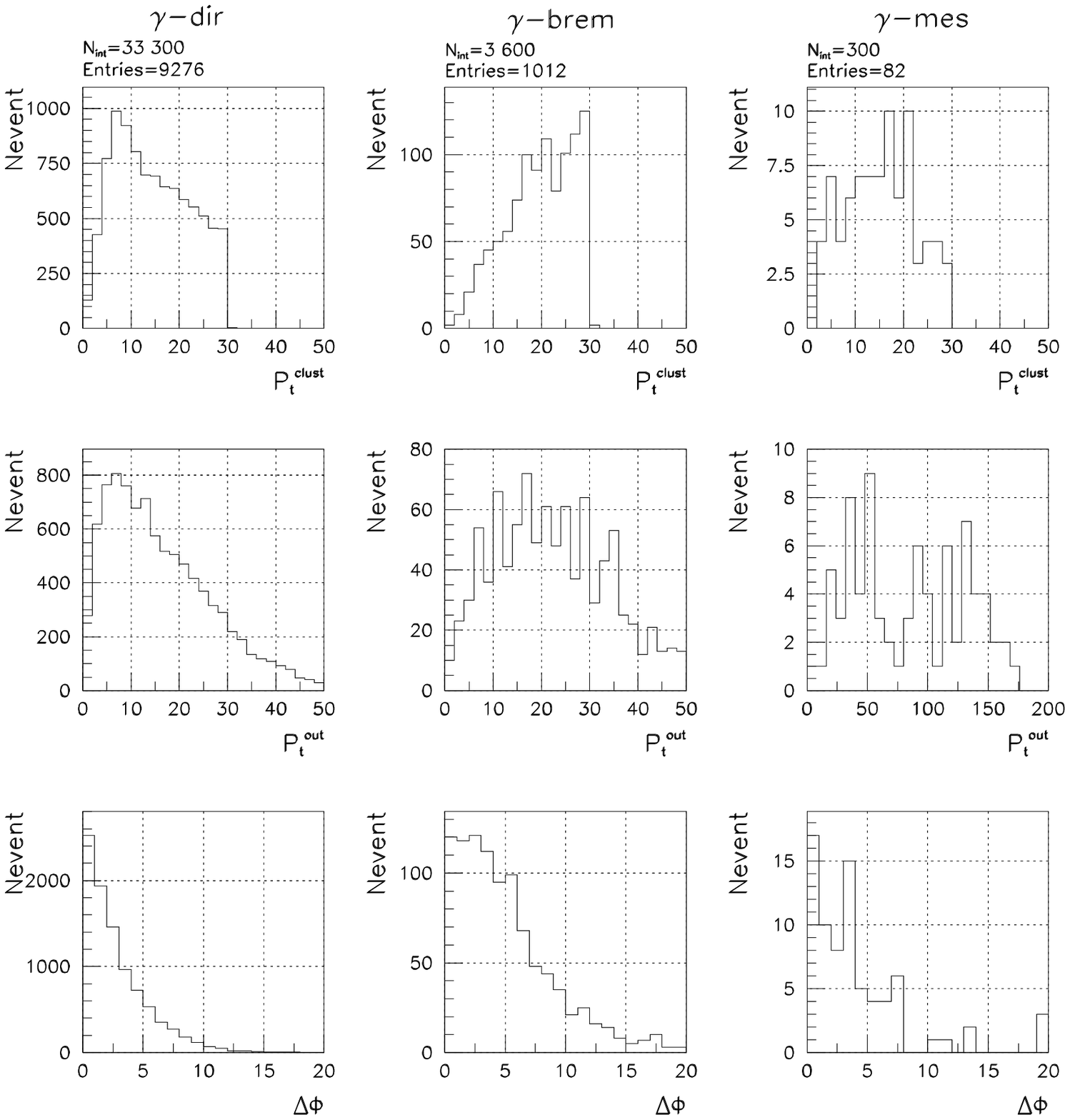}
  \vspace{-0.8cm}
    \caption{\hspace*{0.0cm} \normalsize Signal/Background: Number of events distribution
over $\Pt^{clust}$, $\Pt^{out}$, $\dphi$ ($\Pt^{\gamma}\geq200~GeV/c$)}
    \label{fig:1b200}
  \end{figure}
\begin{figure}[htbp]
 \vspace{0.0cm}
 \hspace*{-0.7cm}
  \includegraphics[width=14cm,height=19cm]{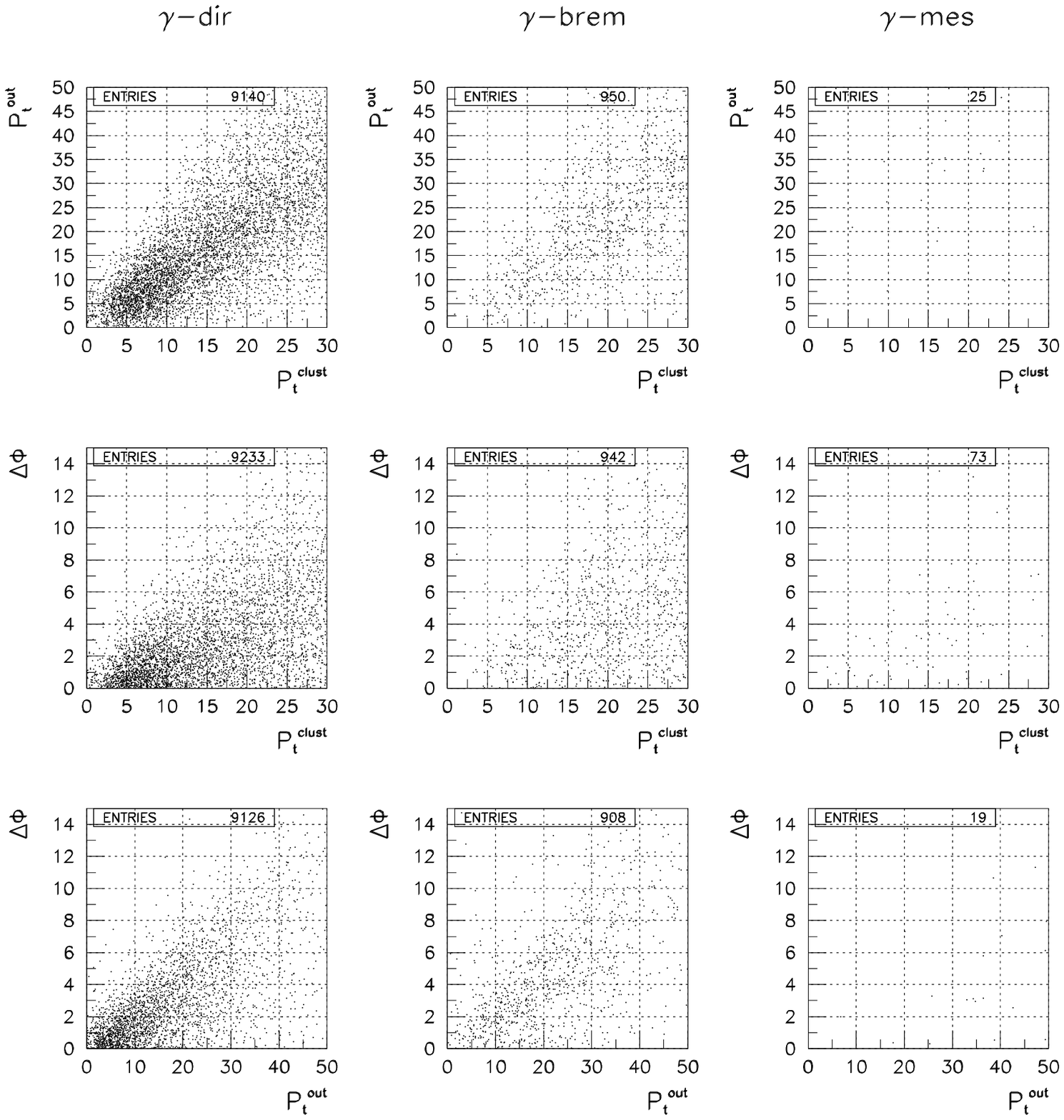}
  \vspace{-0.8cm}
    \caption{\hspace*{0.0cm} \normalsize Signal/Background: $\Pt^{clust}$ vs. $\Pt^{out}$,
$\Pt^{clust}$ vs.$\dphi$, $\Pt^{out}$ vs. $\dphi$ ($\Pt^{\gamma}\geq200~GeV/c$)}
    \label{fig:2b200}
  \end{figure}
\noindent
we conclude that the upper cut $\Delta\phi < 15^{\circ}$
chosen in (26) of [1] is reasonable and indeed 
discards a lot of background events in all $\Pt^{\gamma}$-intervals.

From the second ``$\gamma$-brem'' columns of Figs.~\ref{fig:1b40},
~\ref{fig:1b100} and ~\ref{fig:1b200} one can also
see that $\Pt^{clust}$ spectra of the events with
bremsstrahlung photons look quite different from
the analogous $\Pt^{clust}$ distributions of the signal ''$\gamma$-dir''
photons. The latter distributions have most of the events in the region of
small $\Pt^{clust}$ values.

Since the bremsstrahlung (''$\gamma$-brem'') photons give the most sizable
background, the found difference of the spectra prompts an idea of using
a cut from the top on the value
of $\Pt^{clust}$ to reduce
''$\gamma$-brem'' background which dominates at large $\Pt^{clust}$ values.

The analogous difference of $\Pt^{out}$ spectra of signal
''$\gamma$-dir'' events (which also concentrate at low  $Pt^{out}$ values)
from those of the background ''$\gamma$-brem'' and ''$\gamma$-mes'' events,
with longer tails at high
$\Pt^{out}$ enables us to impose an upper cut on the $\Pt^{out}$ value.

Now from the scatter plots in Figs.~\ref{fig:2b40},
\ref{fig:2b100} and \ref{fig:2b200} as well as from Figs.~\ref{fig:1b40},
\ref{fig:1b100} and \ref{fig:1b200} we can conclude that the usage of cuts
(rather soft here, but their further restriction will be discussed below):
\\[10pt]
\noindent
on $~\Delta \phi$:\\[-9mm]
\begin{center}
$\Delta \phi<15^{\circ},\quad {\rm{for}} \quad Pt^{\gamma}\geq\ 40\;GeV/c;$\\
$\Delta \phi < 10^{\circ},\quad {\rm{for}} \quad Pt^{\gamma}\geq100\;GeV/c;$\\
$\Delta \phi <\ 5^{\circ},\quad {\rm{for}} \quad Pt^{\gamma}\geq200\;GeV/c,$
\end{center}
on $\Pt^{clust}$:\\[-7mm]
\begin{flushleft}
\hspace*{2.1cm} $\Pt^{clust}_{CUT}=15\;GeV/c,\quad {\rm{for}} \quad \Pt^{\gamma}\geq\ 40\; GeV/c$
~and $\;\; \Pt^{\gamma}\geq100\; GeV/c;$\\
\hspace*{2.1cm} $\Pt^{clust}_{CUT}=20\;GeV/c,\quad {\rm{for}} \quad
\Pt^{\gamma}\geq200\; GeV/c,$\\
\end{flushleft}
on $\Pt^{out}_{CUT}$:\\[-4mm]
\vspace{-4mm}
\begin{flushleft}
\hspace*{2.1cm} $\Pt^{out}_{CUT}=10\;GeV/c,\quad {\rm{for}} \quad \Pt^{\gamma}\geq\ 40\; GeV/c;$\\
\hspace*{2.1cm} $\Pt^{out}_{CUT}=15\;GeV/c,\quad {\rm{for}} \quad \Pt^{\gamma}\geq100\; GeV/c$ ~and $\;\; \Pt^{\gamma}\geq200\; GeV/c$
\end{flushleft}
~\\[-2pt]
\noindent
would allow to keep the main part of the signal ''$\gamma$-dir'' events and
to reduce noticeably the contribution from the background ''$\gamma$-brem'' and
''$\gamma$-mes'' events.

So, these figures evidently demonstrate how new physical variables $\Pt^{clust}$ and $\Pt^{out}$,
introduced in Sections 3.1 and 3.2 of paper [1], can be useful for
separation of \gpj events with direct photons from
the background ones (as the latter, in principle, are
not supposed to have good balanced $\Pt^{\gamma}$ and $\Pt^{Jet}$.

\section{DETAILED STUDY OF BACKGROUND SUPPRESSION.}
The study of this Section is based on
the same sample of ``signal + background''  events ($50\cdot10^6$ for each
interval of minimal $\Pt$ of the hard process:~ $\pth=40, ~100, ~200$
$~GeV/c$; see the beginning of Section 2) generated under
the conditions described in the previous section and partially analyzed
there. To demonstrate the way we have come to the final results,
we shall apply the following cuts on the observable
physical variables one after another according to the list presented below.
The influence of these cuts on the $S/B$ ratio is demonstrated in the
following Tables \ref{tab:sb4}--\ref{tab:sb3e}. The numbers in the left-hand
 column
 (``Cut''$\equiv$``Selection'') of  Table  \ref{tab:sb4}, that corresponds to
$\Pt^{\gamma}\geq100 ~GeV/c$, coincide
with the numbers of cuts listed below, where for abbreviation reasons
we shall denote the direct photon by ``$\gamma$'' as well as its candidate.

\begin{table}
\caption{\normalsize List of the applied cuts used in Tables \ref{tab:sb4} -- \ref{tab:sb3}}
\begin{tabular}{lc} \hline
\label{tab:sb0}
{\bf 0}. No cuts; \\
{\bf 1}. $a)~ \Pt^{\gamma}\geq 40 ~GeV/c, ~~b)~|\eta^{\gamma}|\leq 2.61,
~~ c)~ \Pt^{jet}\geq 30 ~GeV/c,~~d) \Pt^{hadr}\!<5 ~GeV/c^{\;\ast}$;\\
{\bf 2}. $\epsilon^{\gamma} \leq 15\%$;
\hspace*{1.87cm} {\bf 12}. $\Pt^{clust}<20 ~GeV/c$; \\
{\bf 3}. $\Pt^{\gamma}\geq\pth$;
\hspace*{1.5cm} {\bf 13}. $\Pt^{clust}<15 ~GeV/c$; \\
{\bf 4}. $\Pt^{sum}\!<1~ GeV/c^{\;\ast\ast}$ ;
\hspace*{0.2cm} {\bf 14}. $\Pt^{clust}<10 ~GeV/c$; \\
{\bf 5}. $\epsilon^{\gamma} \leq 5\%$
\hspace*{2.1cm} {\bf 15}. $\Pt^{out}<20 ~GeV/c$; \\
{\bf 6}. $\Pt^{isol}\!\leq 2~ GeV/c$;
\hspace*{0.76cm}  {\bf 16}. $\Pt^{out}<15 ~GeV/c$; \\
{\bf 7}. $Njet\leq3$;
\hspace*{1.85cm}  {\bf 17}. $\Pt^{out}<\Pt^{out}_{CUT}$\\
{\bf 8}. $Njet\leq2$;
\hspace*{2.4cm}  ($\Pt^{out}_{CUT}=10~ GeV/c$ ~for~ $\pth=100, ~200~ GeV/c$~ \\
{\bf 9}. $Njet=1$;
\hspace*{2.4cm}  and~ $\Pt^{out}_{CUT}=5~GeV/c$ ~for~ $\pth=40~ GeV/c$).\\
{\bf 10}. $\dphi<15^\circ$;
\hspace*{1.63cm} {\bf 18}. $\epsilon^{jet} \leq 5\%$;\\
{\bf 11}. $\Pt^{miss}\!\leq10~GeV/c$;
\hspace*{0.27cm} {\bf 19}. $\epsilon^{jet} \leq 2\%$.\\
\hline
\footnotesize{${\;\ast}~\Pt$ of a hadron in the 5x5 ECAL cell window, containing $\gamma^{dir}$
candidate in the center.}\\
\footnotesize{${\;\ast\ast}$ scalar sum of $\Pt$ in the 5x5 ECAL cell window in the region
out of a smaller 3x3 window, containing $\gamma$.}\\[-3mm]
\end{tabular}
\end{table}
\begin{figure}[h]
\vspace{-1.5cm}
\hspace{.0cm} \includegraphics[width=13cm,height=8cm]{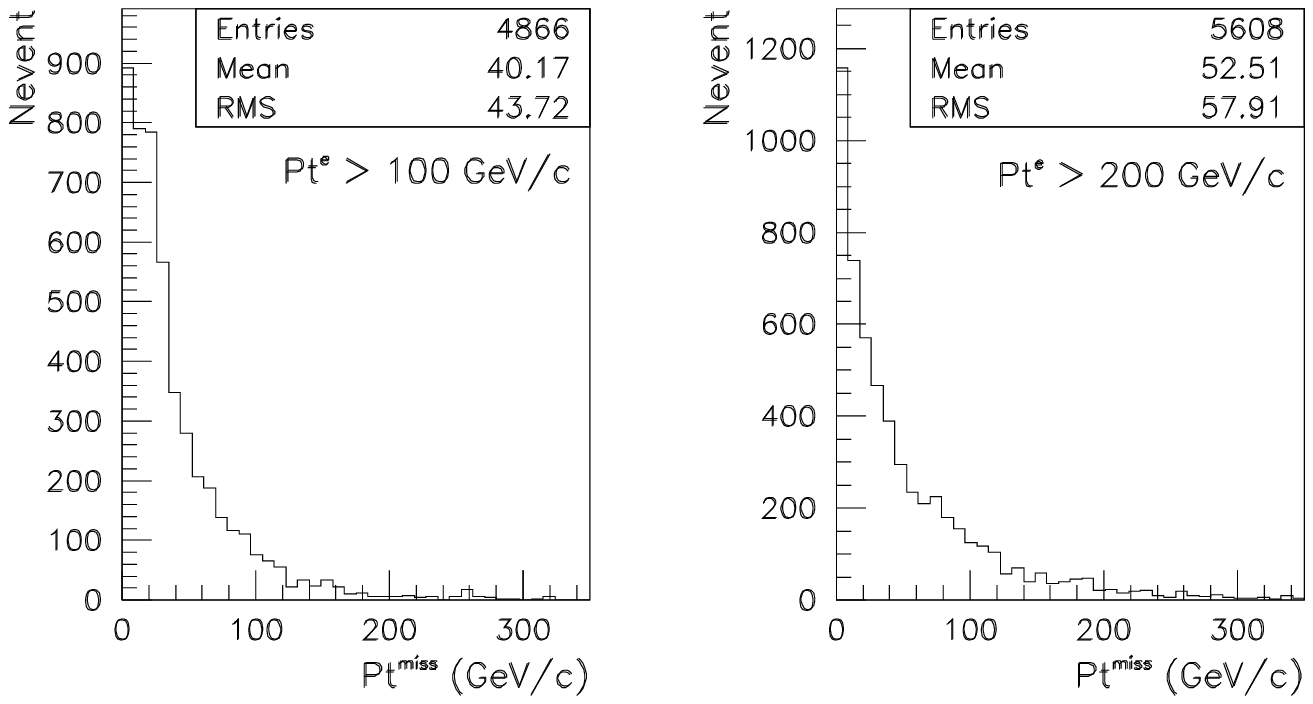}
\vspace{-1.5cm}
\caption{\hspace*{0.0cm}\normalsize Distribution of events over $\Pt^{miss}$ in 
 events with
energetic $e^\pm$`s -- direct photon candidates for the cases ~$\Pt^e\geq100~ GeV/c~$
and $~\Pt^e\geq200~ GeV/c$~ (here are used events satisfying cuts 1--3 of Table 1).}
\label{fig:ptmiss} 
\end{figure}

From the first line of Table \ref{tab:sb4} we see that
without imposing
any cut, the number of the background events exceeds the number of
the signal events, caused by (1a) and (1b) processes, by 5 orders.
The relative isolation cut 2 ($\epsilon^{\gamma} \leq 15\%$) makes the $S/B$ ratio
equal to 0.28.
The cut 3 ($\Pt^{\gamma}\geq\pth$) improves the $S/B$ ratio to 0.68.
The relative isolation cut 5 and then the absolute isolation cut 6 make
the $S/B$ ratio to be equal to 1.50 and 1.93, respectively. The requirement
of only one jet presence in the
event (point 9) results in the value $S/B=5.96$.
The ratio $S/B$ is increased by the cut $\dphi<15^\circ$ up to the value 6.54
(point 10) and at the same time the number of signal events is decreased only by $5\%$.
This is in agreement with the phenomenon of the concentration of events in
the small $\dphi$ angle at large $\Pt^{\gamma}$ values
already mentioned in paper [2].

We have used the $\Pt^{miss}_{CUT}$ cut in paper [1]
to reduce value of $\Pt^{Jet}$ uncertainty due to possible presence of neutrino
contribution to a jet. Here it is applied against the processes based (at the parton
level) on the ~$q\,g \to q' + W^{\pm}$~ and ~$q\bar{~q'} \to g + W^{\pm}$~ subprocesses
with the subsequent decay $W^{\pm} \to e^{\pm}\nu$ and having for this reason
a substantial $\Pt^{miss}$ value. The distributions over $\Pt^{miss}$ for two
$\Pt^e$ values are presented above.
In the last column ($e^\pm$) of Table \ref{tab:sb4} it is shown how
the $\Pt^{miss}_{CUT}$ cut effects the number of these events (point 11).

The reduction of $\Pt^{clust}_{CUT}$ value to $10 ~GeV/c$ (point 14)
results in significant improvement of $S/B$ ratio up to 17.64. Further
reduction of $\Pt^{out}_{CUT}$ value to $10 ~GeV/c$ (point 17) improves
$S/B$ to 22.67. The jet isolation requirement $\epsilon^{jet}<2\%$ (point 19),
finally, gives $S/B=31.05$.\\[-20pt]
\begin{table}[h]
\caption{\normalsize Values of significance and  efficiencies for $\hat{p}_{\perp}^{\;min}$=100 $GeV/c$}
\begin{tabular}{||c||c|c|c|c|c|c||}                  \hline \hline
\label{tab:sb4}
Cut& $S$ & $B^{\;\ast}$ & $Eff_S(\%)$ & $Eff_B(\%)$  & $S/B$& $e^\pm$ \\\hline \hline
 0 & 19420 & 5356.E+6 &             &                 &0.00 &3.9E+6  \\\hline 
 1 & 19359 &  1151425 & 100.00 $\pm$ 0.00& 100.000 $\pm$ 0.000  &0.02 &47061\\\hline 
 2 & 18236 &   65839  & 94.20 $\pm$ 0.97 &   5.718 $\pm$ 0.023  &0.28 &8809 \\\hline 
 3 & 15197 &    22437 &  78.50 $\pm$ 0.85&   1.949 $\pm$ 0.013&  0.68 &2507 \\\hline 
 4 & 15174 &    19005 &  78.38 $\pm$ 0.85&   1.651 $\pm$ 0.012&  0.80 &2486\\\hline 
 5 & 14140 &     9433 &  73.04 $\pm$ 0.81&   0.819 $\pm$ 0.008&  1.50 &2210 \\\hline 
 6 &  8892 &     4618 &  45.93 $\pm$ 0.59&   0.401 $\pm$ 0.006&  1.93 &1331 \\\hline 
 7 & 8572  &    3748  &  44.28 $\pm$ 0.57&   0.326$\pm$  0.005&  2.29 &1174  \\\hline 
 8 & 7663  &    2488  &  39.58 $\pm$ 0.53&   0.216$\pm$  0.004&  3.08 & 921  \\\hline 
 9 & 4844  &     813  &  25.02 $\pm$ 0.40&   0.071$\pm$  0.002&  5.96 & 505  \\\hline 
10 & 4634  &     709  &  23.94 $\pm$ 0.39&   0.062 $\pm$ 0.002&  6.54 & 406 \\\hline 
11 &  4244 &     650  &  21.92 $\pm$ 0.37&   0.056 $\pm$ 0.002&  6.53 &  87 \\\hline 
12 &  3261 &      345 &  16.84 $\pm$ 0.32&   0.030 $\pm$ 0.002&  9.45 &53 \\\hline 
13 &  2558 &      194 &  13.21 $\pm$ 0.28&   0.017 $\pm$ 0.001& 13.19 &41 \\\hline 
14 &  1605 &       91 &   8.29 $\pm$ 0.22&   0.008 $\pm$ 0.001& 17.64 &26 \\\hline 
15 &  1568 &       86 &   8.10 $\pm$ 0.21&   0.007 $\pm$ 0.001& 18.23 &26 \\\hline 
16 &  1477 &       77 &   7.63 $\pm$ 0.21&   0.007 $\pm$ 0.001& 19.18 &25 \\\hline 
17 &  1179 &       52 &   6.09 $\pm$ 0.18&   0.005 $\pm$ 0.001& 22.67 &22 \\\hline 
18 &  1125 &       46 &   5.81 $\pm$ 0.18&   0.004 $\pm$ 0.001& 24.46 &21 \\\hline 
19 &   683 &       22 &   3.53 $\pm$ 0.14&   0.002 $\pm$ 0.000& 31.05 &13 \\\hline 
\hline \hline
\end{tabular}
\end{table}
\vskip-5mm
\noindent
\footnotesize{${\;\ast}$ The background is considered here with no account of contribution
from the `` $e^\pm$ events''}

\vskip2mm
\normalsize
The summary of Table \ref{tab:sb4} is presented in the middle section ($\pth=100 ~GeV/c$)
 of Table \ref{tab:sb1} where in the second column (``Cuts'') line ``Preselected''
corresponds to the 1-st cut of the Table  \ref{tab:sb4} presented above.
 Line ``After cuts'' corresponds
to line 17 and line ``+jet isolation'' --- to the complementary line 19.

In Table \ref{tab:sb1}  the numbers in column ``$\gamma-direct$''
correspond, respectively, to the numbers of the
signal events in column $S$ of Table \ref{tab:sb4} in lines 1, 17 and 19
while the numbers in the 4-th column ``$\gamma-brem$'' correspond to the number
of events with the photons radiated from quarks participating in the hard
 interactions (their $\Pt^{clust}$ and $\Pt^{out}$ distributions
were presented in the central columns of Figs.~1--6 of Section 2).
The columns from 5-th to 8-th of Table \ref{tab:sb1} illustrate
the numbers of the events with the photons ``$\gamma-mes$'',
originated from $\pi^0-,~\eta-,~\omega-,~K^0_S-$ meson decays
(their distributions were shown in the right-hand columns of the same
Figs.~1--6. The total number of the background events without an account of
events with electrons (see last column), i.e.
a sum over columns 4--8 , for the same line
of Table \ref{tab:sb1} is presented, correspondingly, in the column $B$ of
Table \ref{tab:sb4}.\\[-20pt]
\begin{table}[h]
\begin{center}
\caption{\normalsize Number of signal and background events remained after cuts ~~(\bf I)}
\vskip.2cm
\begin{tabular}{||c|c||c|c|c|c|c|c|c||}                  \hline \hline
\label{tab:sb1}
\hmm$\pth$\hmm& &$\gamma$ & $\gamma$ &\multicolumn{4}{c|}{  photons from the mesons}  &
\\\cline{5-8}
\Gvc& Cuts&\hmm direct\hmm &\hmm brem\hmm & $\;\;$ $\pi^0$ $\;\;$ &$\quad$ $\eta$ $\quad$ &
$\omega$ &  $K_S^0$ &\hmm $e^{\pm}$\hmm \\\hline \hline
    &Preselected&\hmm7795&\hmm 12951& 104919& 41845& 10984& 15058&\hmm 4204\hmm  \\\cline{2-9}
 40 &After cuts &\hmm 516&\hmm 48&     41&     10&     0& 5&\hmm   0\hmm\\\cline{2-9}
    &+ jet isol.  &\hmm 122&\hmm 7&      5&     1&     0& 1&   0\\\hline  \hline
    &Preselected&\hmm19359  &\hmm90022 &658981 &247644 &69210  & 85568 &\hmm47061\hmm\\\cline{2-9}
100 &After cuts&\hmm 1179 &\hmm34 &13 &4 &1  & 0 &\hmm 22\hmm \\\cline{2-9}   
    &+ jet isol. &\hmm 683 &\hmm15 &5 &2 &0  & 0 &\hmm 13\hmm \\\hline \hline
    &Preselected&\hmm32629 &\hmm207370 &780190 &288772 &82477 & 98015 &\hmm89714\hmm\\\cline{2-9}
200 &After cuts&\hmm 1074 &\hmm18& 2 &4 &0  & 0 &\hmm 10\hmm\\\cline{2-9}
    &+ jet isol. &\hmm 916 &\hmm16& 1 &4 &0  & 0 &\hmm 8\hmm\\\hline \hline
\end{tabular}
\vskip0.2cm
\caption{\normalsize Efficiencies and significance values in events without jet isolation cut ~~(\bf I)}
\vskip0.1cm
\begin{tabular}{||c||c|c|c|c|c|c||} \hline \hline
\label{tab:sb2}
$\pth$ \Gvc& $S$ & $B$ & $Eff_S(\%)$  & $Eff_B(\%)$  & $S/B$& $S/\sqrt{B}$
\\\hline \hline
40  & 516& 104 & 6.62 $\pm$ 0.30 & 0.056 $\pm$ 0.005&  5.0 & 50.6
 \\\hline
100 & 1179& 52 & 6.09 $\pm$ 0.18 & 0.005 $\pm$ 0.001& 22.7 & 163.5
 \\\hline  
200 & 1074& 24 & 3.29 $\pm$ 0.10 & 0.002 $\pm$ 0.000& 44.8 & 219.2 
\\\hline \hline
\end{tabular}
\vskip0.2cm
\caption{\normalsize Efficiencies and significance values in events with jet isolation cut ~~(\bf I)}
\vskip0.1cm
\begin{tabular}{||c||c|c|c|c|c|c||}  \hline \hline
\label{tab:sb3}
$\pth$ \Gvc& ~~$S$~~ & ~~$B$~~ & $Eff_S(\%)$ & $Eff_B(\%)$  & $S/B$& $S/\sqrt{B}$
 \\\hline \hline
40  & 122& 14 & 1.57 $\pm$ 0.14 & 0.008 $\pm$ 0.002&  8.7 & 32.6 \\\hline 
100 & 683& 22 & 3.53 $\pm$ 0.14 & 0.002 $\pm$ 0.000& 31.1 & 145.6 \\\hline  
200 & 916& 21 & 2.81 $\pm$ 0.09 & 0.001 $\pm$ 0.000& 43.6 & 199.9 
\\\hline \hline
\end{tabular}
\end{center}
\end{table}

\vskip-6mm
The other lines of Table \ref{tab:sb1} for cases $\pth=40$ and
$~200 ~GeV/c~$ have the meaning analogous to those described above
for $\pth=100 ~GeV/c$.

The last column of Table \ref{tab:sb1} shows the number of the events with
electrons which with non-zero
probability can be detected as direct photon.
The following columns of Table \ref{tab:sb4} define efficiencies
$Eff_{S(B)}$ (and their errors) as a ratio
of the number of signal (background) events that passed under some cut
(1--19) to the number of the preselected events (1-st cut of this table).
The last column of Table  \ref{tab:sb4} contains the values of significances
(without account of events with electrons).
The numbers in Tables \ref{tab:sb2} and \ref{tab:sb3}
accumulate in compact form the information of Table \ref{tab:sb1}.
So, for the middle
line ($\pth=100 ~GeV/c$ case) columns $S$ and $B$ contain the numbers of
the signal and background events taken at the level of line 17 (for Table
\ref{tab:sb2}) and line 19 for Table
\ref{tab:sb4}). The column $Eff_{S(B)}$ includes the values of efficiencies
\footnote{taken as a ratio of the number of signal $S$
(background $B$) events, that passed under cut 17 or 19, to the number of
the preselected events in the point 1 of Table \ref{tab:sb4}.}
and their errors.\\[-20pt]
\begin{table}[h]
\begin{center}
\caption{\normalsize Signal vs. background ~~(\bf II)}
\vskip.2cm
\begin{tabular}{||c|c||c|c|c|c|c|c|c||}                  \hline \hline
\label{tab:sb1e}
$\pth$& & $\gamma$ & $\gamma$ &\multicolumn{4}{c|}{  photons from the mesons}  &
\\\cline{5-8}
\Gvc& Cuts&\hmm direct\hmm &\hmm brem\hmm & $\;\;$ $\pi^0$ $\;\;$ &$\quad$ $\eta$ $\quad$ &
$\omega$ &  $K_S^0$ &\hmm $e^{\pm}$\hmm \\\hline \hline
    &Preselected&\hmm 7795&\hmm 12951& 104919& 41845& 10984& 15058&\hmm 4204  \\\cline{2-9}
 40 &After cuts &\hmm 464&\hmm 43&     15&     0&     0&     0&\hmm   0\\\cline{2-9}
    &+ jet isol.  &\hmm 109&\hmm 7&      2&     0&     0&     0&\hmm   0\\\hline  \hline
    &Preselected&\hmm 19359  &\hmm 90022 &658981 &247644 &69210  &85568 &\hmm 47061\\\cline{2-9}
100 &After cuts&\hmm 1061 &\hmm 31 & 9 &0 &0  &0 &\hmm 3 \\\cline{2-9}   
    &+ jet isol. &\hmm 615 &\hmm14 &4 &0 &0  &0 &\hmm 2 \\\hline \hline
    &Preselected&\hmm 32629 &\hmm 207370 &780190 &288772 &82477 &98015 &\hmm 89714\\\cline{2-9}
200 &After cuts&\hmm 967 &\hmm 16& 2 &0 &0  &0 &\hmm 2\\\cline{2-9}
    &+ jet isol. &\hmm 825 &\hmm 14& 1 &0 &0  &0 &\hmm 1\\\hline \hline
\end{tabular}
\vskip0.2cm
\caption{\normalsize Values of efficiencies and significance ~~(\bf II)}
\vskip0.1cm
\begin{tabular}{||c||c|c|c|c|c|c||}   \hline \hline
\label{tab:sb2e}
$\pth$\Gvc& $S$ & $B$ & $Eff_S(\%)$  & $Eff_B(\%)$  & $S/B$& $S/\sqrt{B}$ \\\hline \hline
40  &  464& 58 & 5.95 $\pm$ 0.28 & 0.031 $\pm$ 0.004&  8.0 & 60.9 \\\hline
100 & 1061& 43 & 5.48 $\pm$ 0.17 & 0.004 $\pm$ 0.001& 24.7 & 161.8 \\\hline  
200 & 967& 20 & 2.96 $\pm$ 0.10 & 0.002 $\pm$ 0.000& 48.4 & 216.2 
\\\hline \hline
\end{tabular}
\vskip0.2cm
\caption{\normalsize Values of efficiencies and significance with jet isolation cut ~~(\bf II)}
\vskip0.1cm
\begin{tabular}{||c||c|c|c|c|c|c||}  \hline \hline
\label{tab:sb3e}
$\pth$\Gvc& ~~~$S$~~~ & $B$ & $Eff_S(\%)$ & $Eff_B(\%)$  & $S/B$& $S/\sqrt{B}$
 \\\hline \hline
40  & 109&  9 & 1.40 $\pm$ 0.13 & 0.005 $\pm$ 0.002& 12.1 & 36.3 \\\hline 
100 & 615& 20 & 3.18 $\pm$ 0.13 & 0.003 $\pm$ 0.000& 30.8 & 137.5 \\\hline  
200 & 825& 16 & 2.53 $\pm$ 0.09 & 0.002 $\pm$ 0.000& 51.6 & 206.3 
\\\hline \hline
\end{tabular}
\end{center}
\end{table}

\vskip-6mm
From Table \ref{tab:sb2} it is seen that ratio $S/B$ grows  while
 the \ptg~ value growing from  5.0  at $\Pt^{\gamma}\geq 40 ~GeV/c$ to
44.8 at $\Pt^{\gamma}\geq 200 ~GeV/c$.
The jet isolation requirement (Table \ref{tab:sb3}) sufficiently
improves the situation at low $\Pt$. In that case $S/B$ changes
up to 8.7 at $\Pt^{\gamma}\geq 40 ~GeV/c$ (and up to 31.1
at $\Pt^{\gamma}\geq 100 ~GeV/c$).
Here it is necessary to remind about the conclusion on the
tendency of the selected events to contain an isolated jet:
their number grows while the \ptg~ value increasing.
Practically all jets with $\Pt^{Jet}\geq 200 ~GeV/c$ are isolated
(compare two last lines of Table \ref{tab:sb1})
\footnote{see also Fig.~10 for $\Pt^{\gamma}\geq300 ~GeV/c$ from [1]}.

Up to now we have not used the rejection factors,
that were found basing on the detector capability to discriminate the
background. Let us discuss how
 the values in Tables \ref{tab:sb4}--\ref{tab:sb3}
can be changed by taking into account the real behavior of processes
in the detectors.

We have performed a detailed study (based on CMSIM GEANT simulation using 5000
generated decays of each source meson from Table \ref{tab:sb1})
of difference between ECAL profiles of photon showers
from mesons and those from direct photons for $\Pt^\gamma=40\div 100~ GeV/c$ .
It has shown that the suppression factor of $\eta$-, $\omega$-,
$K_S^0$-mesons larger than 0.90 can be achieved with a selection efficiency of single photons
taken to be 90$\%$. As for the photons from $\pi^0$ decays,~ the analogous~
estimations~ of~ the rejection ~ efficiencies~
were~ done~ for the Endcap~
\begin{figure}[hp]
\vspace{-1.1cm}
\hspace{.0cm} \includegraphics[width=12cm,height=15cm]{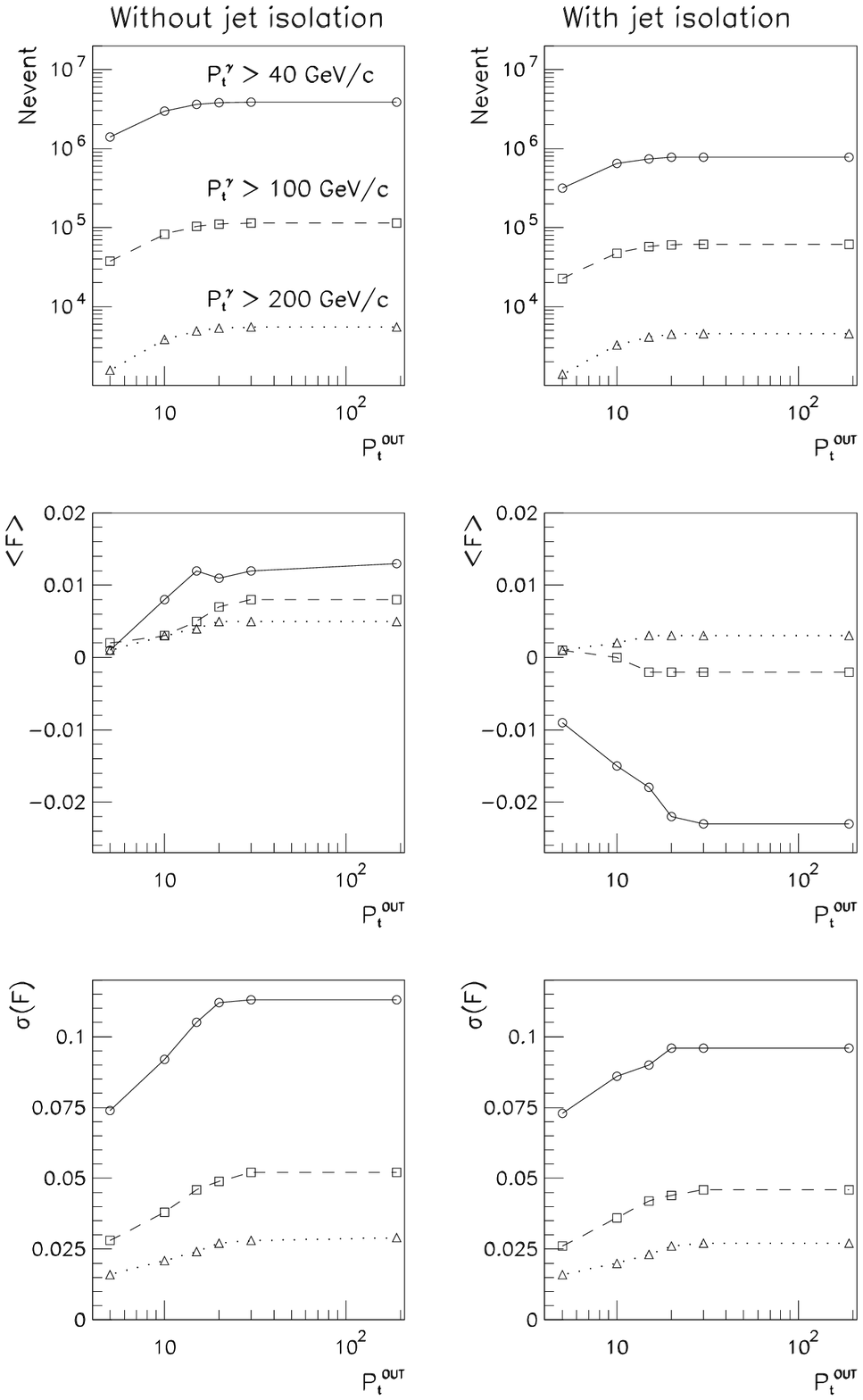}
\vspace{-0.7cm}
\caption{\hspace*{0.0cm}\normalsize The distributions of the number of events
for $L_{int}=3\,fb^{-1}$, the mean value of $(\Pt^{\gamma}-\Pt^{Jet})/\Pt^{\gamma}$
($\equiv\la F\ra$) and its standard deviation $\sigma(F)$ for the cases of
nonisolated (left-hand column) and isolated (right-hand column) jet and for
three \ptg ~ intervals as functions of $\Pt^{out}$ value (in $GeV/c$).
$\Pt^{clust}_{CUT}=10 ~GeV/c.$}
\label{fig:mu-sig}
~\\[0mm]
[6, 8] and Barrel [6, 7] ECAL regions.
They are of the order of 0.20 -- 0.70  for Barrel
and 0.51 -- 0.75 for Endcap, depending on $\Pt^\gamma$ and a bit on
$\eta^{\gamma}$,
for the same single photon selection efficiency 90$\%$.
Following ~[5], for our estimation needs we accept the electron
track finding efficiency to be, on the average, equal to $85\%$ for
$\Pt^e\geq 40 ~GeV/c$, neglecting its $\eta$ dependence.
Then assuming the efficiencies, described above, we have recalculated the numbers
in Tables \ref{tab:sb1}--\ref{tab:sb3}. They are presented in new Tables
\ref{tab:sb1e}--\ref{tab:sb3e} in the analogy with Tables
\ref{tab:sb1}--\ref{tab:sb3}
(the background ($B$) in Tables \ref{tab:sb1e}--\ref{tab:sb3e} differs from one in
Tables \ref{tab:sb1}--\ref{tab:sb3} by including the events with electron candidates
with the corresponding efficiencies).
Comparing new tables with Tables
\ref{tab:sb1}--\ref{tab:sb3} we observe the $50-60\%$ growth of $S/B$ ratio for
$\Pt^\gamma\geq40~GeV/c$ and about the $10\%$ growth for $\Pt^\gamma\geq100~GeV/c$.
\end{figure}

In Tables 3 and 6 we have not presented separately the background due to
$\gamma$/jet misidentification because as it was shown in [9], $\gamma$ and
jet can be discriminated with a high precision and, secondly, as it was mentioned in
Section 2 of this paper (see also Section 3.2 of [1]), we have defined the photon
(or the candidate to be registered as the direct photon) as the signal 
in the 3x3 ECAL
crystal cell window satisfying cut conditions (17) -- (22) of Section 3.2 of [1].
These cuts effectively discriminate the photons from jets.

In Section 2 it has been shown that even with moderate cuts on
$\Pt^{clust}_{CUT}$ and  $\Pt^{out}_{CUT}$ values
the major part of the background events can be suppressed.
A wide variation of these two cuts and their influence on the number of events
(for $L_{int}=3\,fb^{-1}$), and the corresponding values of the
signal to background ratio $S/B$, mean values of vector
disbalance $\Dbgj$, the mean and the standard deviation values for
$(\Pt^{\gamma}\!-\!\Pt^{Jet})/\Pt^{\gamma}$ variable
are presented in Tables \ref{tab:b401} -- \ref{tab:b205} of Appendix.
These tables are built after selections (1) -- (11) of
Table \ref{tab:sb0} in the beginning of
Section 3.  The jet in our 1-jet events
was found by LUCELL jetfinder for the whole $\eta$ region ($|\eta^{jet}|<5.0$).


Tables \ref{tab:b401} -- \ref{tab:b405} correspond to the simulation with
$\pth=40 ~GeV/c$, Tables \ref{tab:b101} -- \ref{tab:b105} --- to
$\pth=100 ~GeV/c$ and  Tables  \ref{tab:b201} -- \ref{tab:b205} --- to
$\pth=200 ~GeV/c$. The  rows and  columns of Tables
\ref{tab:b401} -- \ref{tab:b205} illustrate the influence of
$\Pt^{clust}_{CUT}$ and $\Pt^{out}_{CUT}$ cut values on the quantities
mentioned above, respectively.

All numbers in Tables \ref{tab:b401} -- \ref{tab:b205} are received with account
of the realistic efficiencies of $\gamma^{mes}$ rejection and electron
misidentification as a direct photon.

From Tables \ref{tab:b402}, \ref{tab:b102}, \ref{tab:b202} and
\ref{tab:bi402}, \ref{tab:bi102}, \ref{tab:bi202}
we observe, first of all, noticeable reduction of the background
while moving along the diagonal from the  right-hand bottom corner to the
left-hand upper one, i.e.
with reinforcing $\Pt^{clust}_{CUT}$ and $\Pt^{out}_{CUT}$. So, we see that
for $\pth=40 ~GeV/c$ the ratio $S/B$ changes in the cells along the diagonal
from 3.2, if any limits on these two variables are absent, to 6.5 for
$\Pt^{clust}_{CUT}=10~ GeV/c$ and $\Pt^{out}_{CUT}=10 ~GeV/c$.
Analogously, for $\pth=200 ~GeV/c\,$
the ratio $S/B$ changes from 12.5 to 48.4 with the same variation of
$\Pt^{clust}_{CUT}$ and $\Pt^{out}_{CUT}$.\\
\hspace*{0.8cm} The second observation. The restriction of $\Pt^{clust}$ and $\Pt^{out}$
improves the calibration accuracy. Table
\ref{tab:b404} shows that the mean value of fraction
$F\equiv (\Pt^{\gamma}\!-\!\Pt^{Jet})/\Pt^{\gamma}$
variable decreases from 0.033 (the bottom right-hand corner) to 0.008
for $\Pt^{clust}_{CUT}=10~ GeV/c$ and $\Pt^{out}_{CUT}=10 ~GeV/c$.
Simultaneously (see Tables \ref{tab:b405}, \ref{tab:b105} and
\ref{tab:b205} that include the standard deviation values),
by this restriction one decreases noticeably (about by twice) the width of the gaussian
$\sigma (F)$.

The explanation is simple. The disbalance
equation (29) from [1] contains 2 terms in the right-hand
side: ($1-cos\dphi$) and \Db/$\Pt^{\gamma}$.
The first one is negligibly small and tends to decrease more with
the growth of \ptg~ (see Tables in Appendices of [3]). So,
according to equation (29) of [1],
the main source of the disbalance value is \Db/$\Pt^{\gamma}$ term. While
decreasing $\Pt$ activity out of the jet this term is also decreased and, thus,
the calibration accuracy is increased.

The behavior of the number of events for $L_{int}=3\,fb^{-1}$,
the mean and standard deviation values of the
$(\Pt^{\gamma}\!-\!\Pt^{Jet})/\Pt^{\gamma}$ variable are
also displayed in Fig.~\ref{fig:mu-sig} for isolated and nonisolated jets.

Thus, we can conclude that application of the two criteria introduced
in Section 3.2 from [1], i.e. the cuts on the $\Pt^{clust}$ and $\Pt^{out}$
values, results in two important consequences: significant background reduction
and essential improvement of the calibration accuracy.

In Tables \ref{tab:b403}, \ref{tab:b103}, \ref{tab:b203}
the changes of vector disbalance $\Dbgj$ with variations of the cuts
on the $\Pt^{clust}$ and $\Pt^{out}$ values are presented. The effect is also
evident. The only obvious notice:
the value of $\Dbgj$ in the upper left-hand corner of these Tables (i.e.,
when a $\Pt$ activity out of
a the jet region is almost suppressed), becomes approximately equal to
the value of $\Pt^{\eta>5}$ component (see  line ``$\Pt^{\eta>5}$'' in Appendices of [3]).

The numbers of events for different $\Pt^{clust}_{CUT}$ and $\Pt^{out}_{CUT}$
values are written in the cells of Tables \ref{tab:b401}, \ref{tab:b101} and
\ref{tab:b201}. One can see that even with such strict
$\Pt^{clust}_{CUT}$ and $\Pt^{out}_{CUT}$ values as, i.g. $10 ~GeV/c$ for both,
we would have a sufficient number of events
(3 million, about 80 thousand and 4 thousand for $\Pt^{\gamma}\geq40 ~GeV/c$,
$\Pt^{\gamma}\geq100 ~GeV/c$ and  $\Pt^{\gamma}\geq200 ~GeV/c$, correspondingly)
with low background contamination ($S/B= 6.5,~24.7,~48.4$)
as well as good accuracy of the hadron calorimeter calibration
during one month of continuous LHC running (i.e. $L_{int}= 3~fb^{-1}$)
\footnote{Nevertheless, only full GEANT simulation would allow to come to a
final conclusion.}.

In addition, we also present Tables \ref{tab:bi401}-\ref{tab:bi205}
for the case of the isolated jet by the complete analogy with
Tables \ref{tab:b401}--\ref{tab:b205}.

\normalsize
\def\baselinestretch{1.0}
\section{STUDY OF $\Pt$ BALANCE DEPENDENCE ON PARTON $k_T$.}


This Section is dedicated to the study of possible influence of the intrinsic parton
 transverse momentum on $\Pt$ balance of the \gpj system. For this aim we consider
 two different \\[-20pt]

\begin{table}[h]
\begin{center}
\caption{\normalsize Effect of $k_T$ on $\Pt^{\gamma}$ - $\Pt^{Jet}$ balance with
$\pth\! \!=\!\! 40 ~GeV/c. ~~ F=\Fptgj $}
\vskip0.1cm
\begin{tabular}{||c||c|c|c|c||c|c|c|c||}   \hline  \hline
\label{ap1:tab1}
$\la k_T\ra$& \multicolumn{4}{|c||}{ ISR is OFF}&\multicolumn{4}{c||}{ ISR is
ON} \\\cline{2-9}
$(GeV/c)$ &$\la \Pt56\ra$&$\la \Pt^{5+6}\ra$&$\la F\ra$&$\sigma(F)$
&$\la \Pt56\ra$&$\la \Pt^{5+6}\ra$&$\la F\ra$&$\sigma(F)$   \\\hline  \hline
 0.0 &0.0&0.0 &-0.002 &0.029&8.8 &6.9 &0.007 &0.065\\\hline
 1.0&1.8&1.3 &-0.001 &0.036&9.1 &7.0 &0.009 &0.069  \\\hline
 2.5 &4.5&3.2 &0.001 &0.054 &9.6 &7.4 &0.010 &0.074  \\\hline
 5.0 &8.7&6.1 &0.014 &0.089 &10.4&7.2 &0.015 &0.088  \\\hline
 7.0 &11.2&7.7&0.020 &0.107&11.0 &8.2 &0.022 &0.101  \\\hline  \hline
\end{tabular}
\end{center}
\end{table}

\noindent
ranges of \ptg ~(or $\pth$): $\pth \geq 40~ GeV/c$ and  $\pth \geq 200~ GeV/c$.
For these two $\pth$ values Tables \ref{ap1:tab1} and \ref{ap1:tab2}
demonstrate the average values of $\Pt56$ and $\Pt^{5+6}$ quantities,
defined 

\begin{figure}[t]
\begin{center}
  \includegraphics[width=12cm,height=8.4cm]{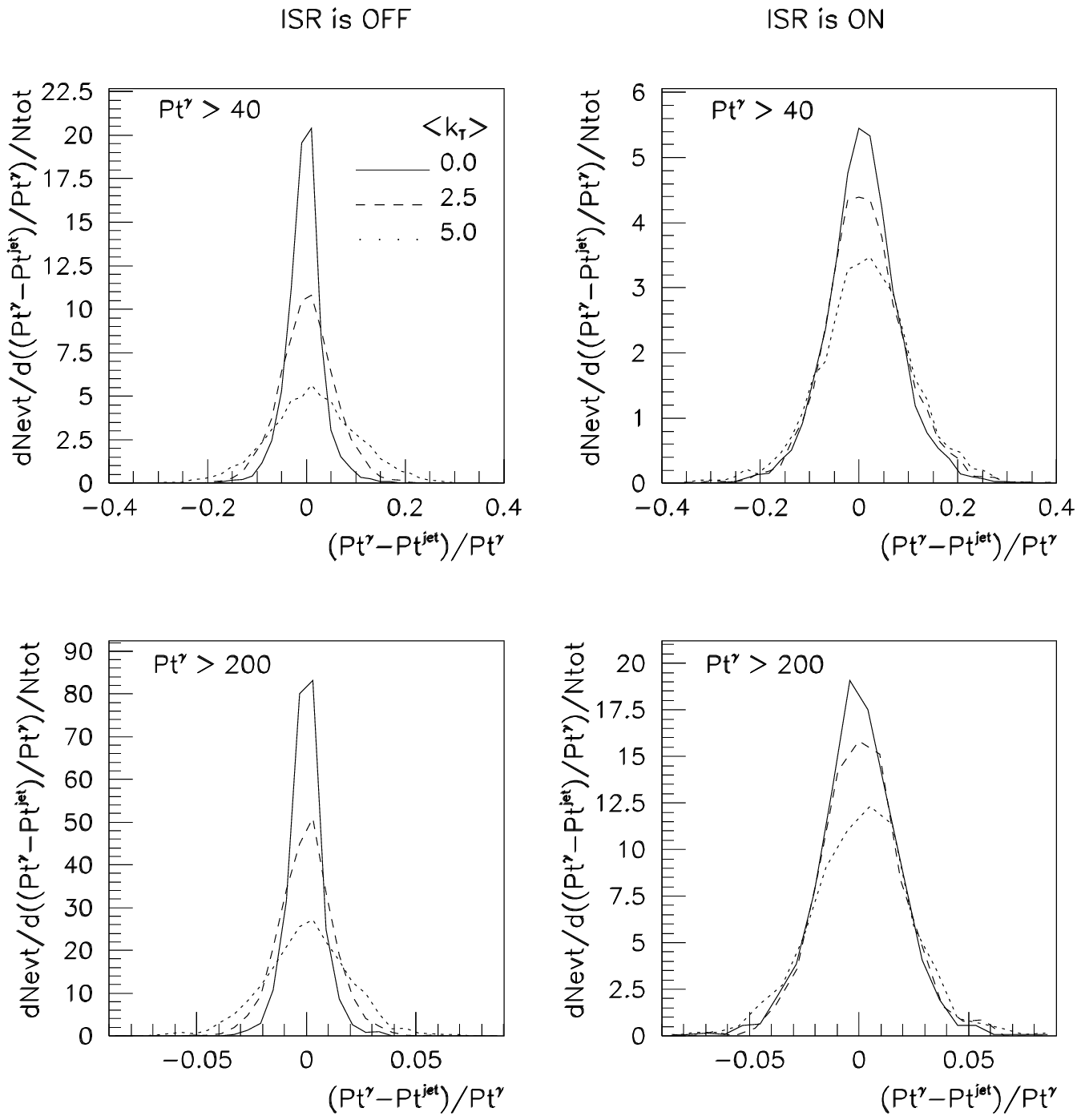}
  \vspace{-0.8cm}
    \caption{\hspace*{0.0cm}\normalsize $\Fptgj$ as a function of primordial $k_T$ value
for cases of switched on and switched off initial radiation and for
$\pth = 40$ and $\pth = 200~~GeV/c$.}
\label{fig:kt}
\end{center}
\end{figure}

\begin{table}[h]
~\\[-14mm]
\caption{\normalsize Effect of $k_T$ on $\Pt^{\gamma}$ -$\Pt^{Jet}$ balance with
$\pth \!\!=\!\! 200~ GeV/c.~ F=\Fptgj $}
\vskip0.0cm
\begin{tabular}[h]{||c||c|c|c|c||c|c|c|c||}                  \hline  \hline
\label{ap1:tab2}
$\la k_T\ra$& \multicolumn{4}{|c||}{ ISR is OFF}&\multicolumn{4}{c||}{ ISR is ON}
\\\cline{2-9}
$(GeV/c)$ &$\la \Pt56\ra$&$\la \Pt^{5+6}\ra$&$\la F\ra$&$\sigma(F)$
&$\la \Pt56\ra$&$\la \Pt^{5+6}\ra$&$\la F\ra$&$\sigma(F)$   \\\hline  \hline
 0.0 &0.0&0.0 &0.000 &0.010 &11.1&8.4 &-0.001 &0.027\\\hline
 1.0 &1.8&1.3 &0.000 &0.013 &11.2&8.6 &0.000  &0.028 \\\hline
 2.5 &4.5&3.1 &0.000 &0.019 &11.8&8.8 &0.001  &0.028 \\\hline
 5.0 &8.7&6.1 &0.001 &0.022 &12.7&9.3 &0.001  &0.031 \\\hline
 7.0 &11.2&7.8 &0.001 &0.029 &13.9&10.4 &0.002 &0.034  \\\hline  \hline
\end{tabular}
~\\[3pt]
\hspace*{0cm}$\ast$ \footnotesize{All numbers in the tables above are given in
 $GeV/c$ units}.
\noindent
\end{table}

\noindent
\normalsize
by (3) of [1], for two different cases: without including
initial state radiation (``ISR is OFF'') 
and with its account (``ISR is ON''),
and various values of parton $\la k_T\ra$: from its absence ($\la k_T\ra=0$)
up to $\la k_T\ra=7 ~GeV/c$. The numbers in Tables
\ref{ap1:tab1} and \ref{ap1:tab2} (obtained from the events set chosen
by the following cuts: $\dphi<15^\circ$,
$\Pt^{out}_{CUT}=5~GeV/c$ and  $\Pt^{clust}_{CUT}=10~ GeV/c$)
 have shown that
the values of $\la\Pt56\ra$ and $\la\Pt{56}\ra$ grow rapidly with
$\la k_T\ra$ increasing if ISR is absent, in fact, with
the values of $\la k_T\ra$. The picture changes when ISR is included.
In this case the disbalance values become large already in the case of
$\la k_T\ra=0$: $\la\Pt56\ra=8.8~ GeV/c$ at $\pth=40 ~GeV/c$
and $\la\Pt^{5+6}\ra=11.1 ~GeV/c$ at 
$\pth=200 ~GeV/c$. The values of
$\la\Pt56\ra$ and $\la\Pt^{5+6}\ra$ grow very slowly from their initial values
at $\la k_T\ra=0$ by $2-3 ~GeV/c$ showing their practical independence on
$\la k_T\ra$ in the range of its reasonable value $\la k_T\ra\leq 1~GeV/c$.

The size of relative disbalance $F=\Fptgj$ variation with $k_T$ is also
shown in Tables \ref{ap1:tab1} and \ref{ap1:tab2} and in plots of
Fig.~\ref{fig:kt}. One can see that for reasonable values $\la k_T\ra\leq 1~GeV/c$
it is quite small.

\section{ ACKNOWLEDGMENTS}                                         

We are greatly thankful to D.~Denegri for having offered this theme to study,
fruitful discussions and permanent support and encouragement.
It is a pleasure for us
to express our recognition for helpful discussions to P.~Aurenche,
M.~Dittmar, M.L.~Mangano,
H.~Rohringer, S.~Taprogge, M.~Fontannaz, J.Ph.~Guillet,
E.~Pilon and J.~Womersley.


\newpage

\begin{table}[h]
{\bf APPENDIX}\\[-8pt]                                             
\def\baselinestretch{0.92}
\begin{center}
\large{ ~$~~\pth= 40~GeV/c$.}
\normalsize
\vskip-0.1cm
\caption{\normalsize Number of events per $L_{int}=3~fb^{-1}$}
\label{tab:b401}
\begin{tabular}{|c||c|c|c|c|c|c|} \hline
$\Pt^{clust}_{\,cut}$ &\multicolumn{6}{c|}{$\Pt^{\,out}_{\,cut}~(GeV/c)$}
\\\cline{2-7}
\Gvc  &\han 5\han&\han 10\han&\han 15\han&\han 20\han&\han 30\han&
 1000  \\\hline \hline
   5&   523000&   885000&   921000&   921000&   921000&   921000\\\hline
   10&  1399000&  2994000&  3642000&  3803000&  3838000&  3841000\\\hline
   15&  1598000&  3809000&  5059000&  5576000&  5745000&  5748000\\\hline
   20&  1678000&  4112000&  5638000&  6408000&  6818000&  6841000\\\hline
   30&  1739000&  4328000&  6082000&  7101000&  7873000&  8101000\\\hline
\end{tabular}
\vskip0.1cm
\caption{ $S/B$}
\label{tab:b402}
\begin{tabular}{|c||c|c|c|c|c|c|} \hline
$\Pt^{clust}_{\,cut}$ &\multicolumn{6}{c|}{$\Pt^{\,out}_{\,cut}~(GeV/c)$}
\\\cline{2-7}
\Gvc  &\has 5\has&\has 10\has&\has 15\has&\has 20\has&\has 30\has&\hass
 1000 \hass \\\hline \hline
    5&  11.6$\pm$ 4.2&   9.2$\pm$ 2.4&   8.8$\pm$ 2.2&   8.8$\pm$ 2.2&   8.8$\pm$ 2.2&   8.8$\pm$ 2.2\\\hline
   10&   8.0$\pm$ 1.5&   6.5$\pm$ 0.8&   5.7$\pm$ 0.6&   5.6$\pm$ 0.6&   5.7$\pm$ 0.6&   5.7$\pm$ 0.6\\\hline
   15&   6.3$\pm$ 1.1&   5.4$\pm$ 0.6&   4.7$\pm$ 0.4&   4.6$\pm$ 0.4&   4.4$\pm$ 0.4&   4.4$\pm$ 0.4\\\hline
   20&   6.1$\pm$ 1.0&   4.8$\pm$ 0.5&   4.1$\pm$ 0.3&   3.9$\pm$ 0.3&   3.8$\pm$ 0.3&   3.8$\pm$ 0.3\\\hline
   30&   5.7$\pm$ 0.9&   4.2$\pm$ 0.4&   3.6$\pm$ 0.3&   3.5$\pm$ 0.2&   3.2$\pm$ 0.2&   3.2$\pm$ 0.2\\\hline
\end{tabular}
\vskip0.1cm
\caption{$\la \Dbgj \ra ~~(GeV/c)$ }
\label{tab:b403}
\begin{tabular}{|c||c|c|c|c|c|c|} \hline
$\Pt^{clust}_{\,cut}$ &\multicolumn{6}{c|}{$\qquad \qquad\Pt^{\,out}_{\,cut}~(GeV/c)\qquad \qquad$}
\\\cline{2-7}
\Gvc&\had 5\had &\had 10\had&\had 15\had&\had 20\had&\had 30\had& \hass 1000 \hass
\\\hline\hline
    5&      3.8&      4.5&      4.7&      4.7&      4.7&      4.7\\\hline
   10&      4.2&      5.4&      6.2&      6.6&      6.7&      6.7\\\hline
   15&      4.2&      5.7&      7.0&      7.7&      8.1&      8.1\\\hline
   20&      4.3&      5.8&      7.2&      8.3&      9.1&      9.1\\\hline
   30&      4.3&      5.9&      7.5&      8.7&     10.0&     10.8\\\hline
\end{tabular}
\vskip0.1cm
\caption{$\la F\ra,~F= \Fptgj$ }
\label{tab:b404}
\begin{tabular}{|c||c|c|c|c|c|c|} \hline
$\Pt^{clust}_{\,cut}$ &\multicolumn{6}{c|}{$\qquad \qquad\Pt^{\,out}_{\,cut}~(GeV/c)\qquad \qquad$}
\\\cline{2-7}
\Gvc&\had 5\had &\had 10\had&\had 15\had&\had 20\had&\had 30\had& \hass 1000 \hass\\\hline\hline
    5&    0.005&    0.006&    0.007&    0.007&    0.007&    0.007\\\hline
   10&    0.001&    0.008&    0.012&    0.011&    0.012&    0.013\\\hline
   15&    0.004&    0.010&    0.017&    0.019&    0.022&    0.022\\\hline
   20&    0.005&    0.011&    0.021&    0.024&    0.029&    0.029\\\hline
   30&    0.004&    0.011&    0.022&    0.026&    0.032&    0.033\\\hline
\end{tabular}
\vskip0.1cm
\caption{$\sigma(F),~F= \Fptgj$ }
\label{tab:b405}
\begin{tabular}{|c||c|c|c|c|c|c|} \hline
$\Pt^{clust}_{\,cut}$ &\multicolumn{6}{c|}{$\qquad \qquad\Pt^{\,out}_{\,cut}~(GeV/c)\qquad \qquad$}
\\\cline{2-7}
\Gvc&\had 5\had &\had 10\had&\had 15\had&\had 20\had&\had 30\had& \hass 1000 \hass\\\hline\hline
    5&    0.069&    0.081&    0.085&    0.085&    0.085&    0.085\\\hline
   10&    0.074&    0.092&    0.105&    0.112&    0.113&    0.113\\\hline
   15&    0.076&    0.098&    0.118&    0.134&    0.140&    0.141\\\hline
   20&    0.076&    0.100&    0.123&    0.144&    0.158&    0.159\\\hline
   30&    0.077&    0.102&    0.127&    0.152&    0.173&    0.177\\\hline
\end{tabular}
\end{center}
\end{table}
\def\baselinestretch{0.95}
\begin{table}[htbp]
\begin{center}
\large{ ~$~~\pth= 100~GeV/c$.}
\normalsize
\caption{\normalsize Number of events per $L_{int}=3~fb^{-1}$}
\label{tab:b101}
\begin{tabular}{|c||c|c|c|c|c|c|} \hline
$\Pt^{clust}_{\,cut}$ &\multicolumn{6}{c|}{$\Pt^{\,out}_{\,cut}~(GeV/c)$}
\\\cline{2-7}
\Gvc&\hbn 5\hbn &\hbn 10\hbn&\hbn 15\hbn&\hbn 20\hbn&\hbn 30\hbn&
 1000  \\\hline\hline
    5&    14100&    24600&    27000&    27200&    27300&    27400\\\hline
   10&    37700&    82400&   103800&   110600&   113500&   113600\\\hline
   15&    44900&   106300&   146800&   168500&   183300&   184200\\\hline
   20&    47100&   114600&   166600&   200900&   234300&   241400\\\hline
   30&    48900&   121200&   180600&   227900&   293900&   327200\\\hline
\end{tabular}
\vskip0.1cm
\caption{$S/B $}
\label{tab:b102}
\begin{tabular}{|c||c|c|c|c|c|c|} \hline
$\Pt^{clust}_{\,cut}$ &\multicolumn{6}{c|}{$\Pt^{\,out}_{\,cut}~(GeV/c)$}
\\\cline{2-7}
\Gvc&\hn 5\hn&\hn 10\hn&\hn 15\hn&\hn 20\hn&\hn 30\hn&\hn 1000\hn \\\hline \hline
    5&\hm  59.0$\pm$38.2&\hm  44.3$\pm$19.2&\hm  40.7$\pm$16.2&\hm  41.0$\pm$16.4&\hm  41.0$\pm$16.4&\hm  41.0$\pm$16.4\\\hline
   10&\hm  25.0$\pm$ 6.5&\hm  24.7$\pm$ 4.6&\hm  21.4$\pm$ 3.3&\hm  20.5$\pm$ 3.0&\hm  19.5$\pm$ 2.8&\hm  19.8$\pm$ 2.8\\\hline
   15&\hm  23.9$\pm$ 6.0&\hm  19.3$\pm$ 2.9&\hm  16.1$\pm$ 1.9&\hm 15.2$\pm$ 1.6&\hm  14.1$\pm$ 1.4&\hm  14.1$\pm$ 1.4\\\hline
   20&\hm  19.6$\pm$ 4.4&\hm  15.9$\pm$ 2.1&\hm  12.8$\pm$ 1.3&\hm  12.0$\pm$ 1.1&\hm  10.1$\pm$ 0.8&\hm   9.9$\pm$ 0.8\\\hline
   30&\hm  18.6$\pm$ 4.0&\hm  13.6$\pm$ 1.7&\hm  11.0$\pm$ 1.0&\hm   9.2$\pm$ 0.7&\hm   7.4$\pm$ 0.5&\hm   6.8$\pm$ 0.4\\\hline
\end{tabular}
\vskip0.1cm
\caption{$\la \Dbgj \ra ~~(GeV/c)$}
\label{tab:b103}
\begin{tabular}{|c||c|c|c|c|c|c|} \hline
$\Pt^{clust}_{\,cut}$ &\multicolumn{6}{c|}{$\qquad \qquad\Pt^{\,out}_{\,cut}~(GeV/c)\qquad \qquad$}
\\\cline{2-7}
\Gvc&\hbd 5\hbd &\hbd 10\hbd&\hbd 15\hbd&\hbd 20\hbd&\hbd 30\hbd&
\hbd 1000 \hbd \\\hline\hline
    5&      3.7&      4.7&      5.2&      5.3&      5.4&      5.5\\\hline
   10&      4.0&      5.5&      6.6&      7.2&      7.5&      7.6\\\hline
   15&      4.2&      5.9&      7.5&      8.6&      9.7&      9.8\\\hline
   20&      4.3&      6.1&      7.9&      9.4&     11.3&     11.9\\\hline
   30&      4.4&      6.1&      8.1&     10.0&     13.0&     15.5\\\hline
\end{tabular}
\vskip0.1cm
\caption{$\la F\ra, ~F= \Fptgj$}
\label{tab:b104}
\begin{tabular}{|c||c|c|c|c|c|c|} \hline
$\Pt^{clust}_{\,cut}$ &\multicolumn{6}{c|}{$\qquad \qquad\Pt^{\,out}_{\,cut}~(GeV/c)\qquad \qquad$}
\\\cline{2-7}
\Gvc&\hbd 5\hbd &\hbd 10\hbd&\hbd 15\hbd&\hbd 20\hbd&\hbd 30\hbd&
\hbd 1000 \hbd \\\hline\hline
    5&    0.002&    0.001&    0.003&    0.004&    0.004&    0.005\\\hline
   10&    0.002&    0.003&    0.005&    0.007&    0.008&    0.008\\\hline
   15&    0.001&    0.003&    0.006&    0.009&    0.013&    0.013\\\hline
   20&    0.002&    0.003&    0.006&    0.011&    0.016&    0.019\\\hline
   30&    0.002&    0.003&    0.005&    0.011&    0.019&    0.027\\\hline
\end{tabular}
\vskip0.1cm
\caption{$\sigma(F),~F= \Fptgj$ }
\label{tab:b105}
\begin{tabular}{|c||c|c|c|c|c|c|} \hline
$\Pt^{clust}_{\,cut}$ &\multicolumn{6}{c|}{$\qquad \qquad\Pt^{\,out}_{\,cut}~(GeV/c)\qquad \qquad$}
\\\cline{2-7}
\Gvc&\hbd 5\hbd &\hbd 10\hbd&\hbd 15\hbd&\hbd 20\hbd&\hbd 30\hbd&
\hbd 1000 \hbd \\\hline\hline
    5&    0.028&    0.035&    0.038&    0.038&    0.039&    0.039\\\hline
   10&    0.028&    0.038&    0.046&    0.049&    0.052&    0.052\\\hline
   15&    0.029&    0.041&    0.051&    0.057&    0.065&    0.066\\\hline
   20&    0.030&    0.042&    0.053&    0.062&    0.075&    0.080\\\hline
   30&    0.030&    0.043&    0.055&    0.067&    0.087&    0.101\\\hline
\end{tabular}
\end{center}
\end{table}
\def\baselinestretch{0.95}
\begin{table}[htbp]
\begin{center}
\large{ ~$~~\pth= 200~GeV/c$.}
\normalsize
\caption{\normalsize Number of events per $L_{int}=3~fb^{-1}$}
\label{tab:b201}
\vskip-0.0cm
\begin{tabular}{|c||c|c|c|c|c|c|} \hline
$\Pt^{clust}_{\,cut}$ &\multicolumn{6}{c|}{$\Pt^{\,out}_{\,cut}~(GeV/c)$}
\\\cline{2-7}
\Gvc  &\hcn 5\hcn&\hcn 10\hcn&\hcn 15&\hcn 20\hcn&\hcn 30\hcn&\hcn 1000\hcn \\\hline\hline
    5&      570&     1090&     1180&     1210&     1220&     1230\\\hline
   10&     1550&     3830&     4900&     5360&     5470&     5490\\\hline
   15&     1960&     5060&     7360&     8710&     9500&     9630\\\hline
   20&     2110&     5590&     8480&    10550&    12410&    12990\\\hline
   30&     2170&     5880&     9210&    11830&    15510&    18270\\\hline
\end{tabular}
\vskip0.1cm
\caption{$S/B$}
\label{tab:b202}
\begin{tabular}{|c||c|c|c|c|c|c|} \hline
$\Pt^{clust}_{\,cut}$ &\multicolumn{6}{c|}{$\Pt^{\,out}_{\,cut}~(GeV/c)$}
\\\cline{2-7}
\Gvc  & 5& 10& 15& 20& 30& 1000  \\\hline \hline
    5& 109 $\pm$ 109&173 $\pm$ 154& 115 $\pm$ 82& 118 $\pm$ 84&118 $\pm$ 84&118 $\pm$ 84\\\hline
   10&  50.7$\pm$16.0&  48.4$\pm$13.3&  44.1$\pm$10.2&  37.4$\pm$ 7.6&  38.2$\pm$ 7.8&  38.2$\pm$ 7.8\\\hline
   15&  38.5$\pm$13.2&  44.1$\pm$10.0&  35.9$\pm$ 6.2&  27.9$\pm$ 4.0&  25.7$\pm$ 3.4&  25.4$\pm$ 3.3\\\hline
   20&  29.3$\pm$ 8.6&  36.4$\pm$ 7.2&  26.5$\pm$ 3.7&  22.0$\pm$ 2.6&  18.8$\pm$ 1.9&  17.2$\pm$ 1.7\\\hline
   30&  28.6$\pm$ 8.2&  28.2$\pm$ 4.9&  20.6$\pm$ 2.5&  17.4$\pm$ 1.8&  14.5$\pm$ 1.2&  12.5$\pm$ 0.9\\\hline
\end{tabular}
\vskip0.1cm
\caption{$\la \Dbgj \ra ~~(GeV/c)$}
\label{tab:b203}
\begin{tabular}{|c||c|c|c|c|c|c|} \hline\
$\Pt^{clust}_{\,cut}$ &\multicolumn{6}{c|}{$\qquad \qquad\Pt^{\,out}_{\,cut}~(GeV/c)\qquad \qquad$}
\\\cline{2-7}
\Gvc  &\hcd 5\hcd&\hcd 10\hcd&\hcd 15\hcd&\hcd 20\hcd&\hcd 30\hcd&
\hcd 1000 \hcd \\\hline \hline
    5&      4.0&      5.0&      5.4&      5.6&      5.7&      6.1\\\hline
   10&      4.5&      6.0&      7.0&      7.8&      8.1&      8.2\\\hline
   15&      4.6&      6.2&      7.9&      9.2&     10.4&     10.7\\\hline
   20&      4.7&      6.3&      8.2&      9.9&     11.9&     13.0\\\hline
   30&      4.7&      6.4&      8.5&     10.4&     13.7&     17.5\\\hline
\end{tabular}
\vskip0.1cm
\caption{$\la F\ra, ~F=\Fptgj$}
\label{tab:b204}
\begin{tabular}{|c||c|c|c|c|c|c|} \hline
$\Pt^{clust}_{\,cut}$ &\multicolumn{6}{c|}{$\qquad \qquad\Pt^{\,out}_{\,cut}~(GeV/c)\qquad \qquad$}
\\\cline{2-7}
\Gvc  &\hcd 5\hcd&\hcd 10\hcd&\hcd 15\hcd&\hcd 20\hcd&\hcd 30\hcd&
\hcd 1000 \hcd \\\hline \hline
    5&    0.003&    0.003&    0.003&    0.004&    0.004&    0.005\\\hline
   10&    0.001&    0.003&    0.004&    0.005&    0.005&    0.005\\\hline
   15&    0.001&    0.003&    0.006&    0.007&    0.008&    0.008\\\hline
   20&    0.001&    0.003&    0.005&    0.007&    0.008&    0.010\\\hline
   30&    0.001&    0.003&    0.005&    0.007&    0.009&    0.014\\\hline
\end{tabular}
\vskip0.1cm
\caption{$\sigma(F),~F= \Fptgj$ }
\label{tab:b205}
\begin{tabular}{|c||c|c|c|c|c|c|} \hline
$\Pt^{clust}_{\,cut}$ &\multicolumn{6}{c|}{$\qquad \qquad\Pt^{\,out}_{\,cut}~(GeV/c)\qquad \qquad$}
\\\cline{2-7}
\Gvc  &\hcd 5\hcd&\hcd 10\hcd&\hcd 15\hcd&\hcd 20\hcd&\hcd 30\hcd&
\hcd 1000 \hcd \\\hline
    5&    0.015&    0.018&    0.020&    0.021&    0.023&    0.026\\\hline
   10&    0.016&    0.021&    0.024&    0.027&    0.028&    0.029\\\hline
   15&    0.016&    0.021&    0.027&    0.031&    0.035&    0.037\\\hline
   20&    0.016&    0.022&    0.028&    0.033&    0.040&    0.044\\\hline
   30&    0.016&    0.022&    0.029&    0.035&    0.046&    0.057\\\hline
\end{tabular}
\end{center}
\end{table}


\def\baselinestretch{0.95}
\begin{table}[htbp]
\begin{center}
\large{ ~$~~\pth= 40~GeV/c$,  ~~~~$\epsilon^{jet}<2\%$.}
\normalsize
\caption{\normalsize Number of events per $L_{int}=3~fb^{-1}$}
\label{tab:bi401}
\begin{tabular}{|c||c|c|c|c|c|c|} \hline
$\Pt^{clust}_{\,cut}$ &\multicolumn{6}{c|}{$\Pt^{\,out}_{\,cut}~(GeV/c)$}
\\\cline{2-7}
\Gvc  &\han 5\han&\han 10\han&\han 15\han&\han 20\han&\han 30\han&
 1000  \\\hline \hline
    5&   165000&   278000&   284000&   284000&   284000&   284000\\\hline
   10&   317000&   652000&   739000&   772000&   778000&   778000\\\hline
   15&   332000&   737000&   890000&   970000&   994000&   994000\\\hline
   20&   355000&   779000&   958000&  1074000&  1124000&  1124000\\\hline
   30&   361000&   805000&  1000000&  1146000&  1266000&  1308000\\\hline
\end{tabular}
\vskip0.1cm
\caption{ $S/B$}
\label{tab:bi402}
\begin{tabular}{|c||c|c|c|c|c|c|} \hline 
$\Pt^{clust}_{\,cut}$ &\multicolumn{6}{c|}{$\Pt^{\,out}_{\,cut}~(GeV/c)$}
\\\cline{2-7}
\Gvc  &\has 5\has&\has 10\has&\has 15\has&\has 20\has&\has 30\has&\hass
 1000 \hass \\\hline \hline
    5&   8.6$\pm$ 5.0&  12.6$\pm$ 6.5&  12.3$\pm$ 6.2&  12.3$\pm$ 6.2&  12.3$\pm$ 6.2&  12.3$\pm$ 6.2\\\hline
   10&  12.1$\pm$ 5.7&   9.6$\pm$ 3.0&   8.2$\pm$ 2.1&   7.8$\pm$ 2.0&   7.8$\pm$ 2.0&   7.8$\pm$ 2.0\\\hline
   15&  11.2$\pm$ 5.0&   8.8$\pm$ 2.5&   7.1$\pm$ 1.7&   6.6$\pm$ 1.5&   6.3$\pm$ 1.4&   6.4$\pm$ 1.4\\\hline
   20&  10.8$\pm$ 4.6&   7.9$\pm$ 2.1&   6.7$\pm$ 1.5&   6.2$\pm$ 1.3&   5.9$\pm$ 1.2&   6.0$\pm$ 1.2\\\hline
   30&  11.0$\pm$ 4.7&   7.1$\pm$ 1.8&   6.0$\pm$ 1.3&   5.6$\pm$ 1.1&   5.2$\pm$ 0.9&   4.9$\pm$ 0.9\\\hline
\end{tabular}
\vskip0.1cm
\caption{$\la \Dbgj \ra ~~(GeV/c)$ }
\label{tab:bi403}
\begin{tabular}{|c||c|c|c|c|c|c|} \hline
$\Pt^{clust}_{\,cut}$ &\multicolumn{6}{c|}{$\qquad \qquad\Pt^{\,out}_{\,cut}~(GeV/c)\qquad \qquad$}
\\\cline{2-7}
\Gvc&\had 5\had &\had 10\had&\had 15\had&\had 20\had&\had 30\had& \hass 1000 \hass
\\\hline\hline
    5&      4.1&      4.5&      4.7&      4.7&      4.7&      4.7\\\hline
   10&      4.5&      5.4&      6.0&      6.2&      6.4&      6.4\\\hline
   15&      4.4&      5.5&      6.4&      7.1&      7.4&      7.4\\\hline
   20&      4.3&      5.5&      6.4&      7.4&      8.0&      8.0\\\hline
   30&      4.3&      5.5&      6.6&      7.7&      9.2&     10.0\\\hline
\end{tabular}
\vskip0.1cm
\caption{$\la F\ra,~F= \Fptgj$ }
\label{tab:bi404}
\begin{tabular}{|c||c|c|c|c|c|c|} \hline
$\Pt^{clust}_{\,cut}$ &\multicolumn{6}{c|}{$\qquad \qquad\Pt^{\,out}_{\,cut}~(GeV/c)\qquad \qquad$}
\\\cline{2-7}
\Gvc&\had 5\had &\had 10\had&\had 15\had&\had 20\had&\had 30\had& \hass 1000 \hass\\\hline\hline
    5&   -0.004&   -0.009&   -0.007&   -0.007&   -0.007&   -0.007\\\hline
   10&   -0.009&   -0.015&   -0.018&   -0.022&   -0.023&   -0.023\\\hline
   15&   -0.007&   -0.013&   -0.015&   -0.021&   -0.027&   -0.027\\\hline
   20&   -0.006&   -0.011&   -0.013&   -0.017&   -0.021&   -0.021\\\hline
   30&   -0.006&   -0.012&   -0.012&   -0.016&   -0.025&   -0.033\\\hline
\end{tabular}
\vskip0.1cm
\caption{$\sigma(F),~F= \Fptgj$ }
\label{tab:bi405}
\begin{tabular}{|c||c|c|c|c|c|c|} \hline
$\Pt^{clust}_{\,cut}$ &\multicolumn{6}{c|}{$\qquad \qquad\Pt^{\,out}_{\,cut}~(GeV/c)\qquad \qquad$}
\\\cline{2-7}
\Gvc&\had 5\had &\had 10\had&\had 15\had&\had 20\had&\had 30\had& \hass 1000 \hass\\\hline\hline
    5&    0.073&    0.080&    0.082&    0.082&    0.082&    0.082\\\hline
   10&    0.073&    0.086&    0.090&    0.096&    0.096&    0.096\\\hline
   15&    0.074&    0.088&    0.098&    0.115&    0.119&    0.119\\\hline
   20&    0.073&    0.087&    0.098&    0.120&    0.131&    0.131\\\hline
   30&    0.072&    0.089&    0.102&    0.129&    0.153&    0.158\\\hline
\end{tabular}
\end{center}
\end{table}

\def\baselinestretch{0.95}
\begin{table}[htbp]
\begin{center}
\large{ ~$~~\pth= 100~GeV/c$,  ~~~~$\epsilon^{jet}<2\%$.}
\normalsize
\caption{\normalsize Number of events per $L_{int}=3~fb^{-1}$}
\label{tab:bi101}
\begin{tabular}{|c||c|c|c|c|c|c|} \hline 
$\Pt^{clust}_{\,cut}$ &\multicolumn{6}{c|}{$\Pt^{\,out}_{\,cut}~(GeV/c)$}
\\\cline{2-7}
\Gvc&\hbn 5\hbn &\hbn 10\hbn&\hbn 15\hbn&\hbn 20\hbn&\hbn 30\hbn&
 1000  \\\hline\hline
    5&     9900&    15600&    16900&    17000&    17000&    17000\\\hline
   10&    22700&    47400&    57700&    60400&    61500&    61500\\\hline
   15&    26600&    59500&    77700&    86400&    92100&    92600\\\hline
   20&    27800&    63100&    86100&    99700&   112600&   114500\\\hline
   30&    28300&    65500&    91600&   108700&   134200&   144300\\\hline
\end{tabular}
\vskip0.1cm
\caption{$S/B $}
\label{tab:bi102}
\begin{tabular}{|c||c|c|c|c|c|c|} \hline 
$\Pt^{clust}_{\,cut}$ &\multicolumn{6}{c|}{$\Pt^{\,out}_{\,cut}~(GeV/c)$}
\\\cline{2-7}
\Gvc & 5& 10& 15& 20& 30&  1000 \\\hline \hline
    5&\hm  58.6$\pm$45.2&\hm  56.8$\pm$34.4& \hm 55.8$\pm$32.2&\hm  56.0$\pm$32.3&\hm  56.0$\pm$32.3&\hm  56.0$\pm$32.3\\\hline
   10&\hm  30.6$\pm$11.8&\hm  30.8$\pm$ 8.4&\hm  26.7$\pm$ 5.7&\hm  27.9$\pm$ 6.0&\hm  27.3$\pm$ 5.8&\hm  27.3$\pm$ 5.8\\\hline
   15& \hm 32.8$\pm$12.0&\hm  26.5$\pm$ 6.0&\hm  21.2$\pm$ 3.8&\hm  22.2$\pm$ 3.9&\hm  20.8$\pm$ 3.4&\hm  21.1$\pm$ 3.4\\\hline
   20&\hm  29.0$\pm$ 9.9&\hm  21.6$\pm$ 4.3&\hm  17.2$\pm$ 2.7&\hm  17.8$\pm$ 2.7&\hm  15.5$\pm$ 2.1&\hm  15.7$\pm$ 2.1\\\hline
   30&\hm  27.4$\pm$ 9.1&\hm  19.7$\pm$ 3.8&\hm  16.2$\pm$ 2.4&\hm  15.4$\pm$ 2.1&\hm  12.2$\pm$ 1.4&\hm  11.5$\pm$ 1.2\\\hline
\end{tabular}
\vskip0.1cm
\caption{$\la \Dbgj \ra ~~(GeV/c)$}
\label{tab:bi103}
\begin{tabular}{|c||c|c|c|c|c|c|} \hline
$\Pt^{clust}_{\,cut}$ &\multicolumn{6}{c|}{$\qquad \qquad\Pt^{\,out}_{\,cut}~(GeV/c)\qquad \qquad$}
\\\cline{2-7}
\Gvc&\hbd 5\hbd &\hbd 10\hbd&\hbd 15\hbd&\hbd 20\hbd&\hbd 30\hbd&
\hbd 1000 \hbd \\\hline\hline
    5&      3.5&      4.5&      5.0&      5.0&      5.0&      5.0\\\hline
   10&      3.9&      5.3&      6.4&      6.8&      7.0&      7.0\\\hline
   15&      4.1&      5.7&      7.1&      8.1&      8.9&      9.0\\\hline
   20&      4.3&      5.9&      7.5&      8.7&     10.2&     10.6\\\hline
   30&      4.3&      5.9&      7.7&      9.1&     11.8&     13.4\\\hline
\end{tabular}
\vskip0.1cm
\caption{$\la F\ra, ~F= \Fptgj$}
\label{tab:bi104}
\begin{tabular}{|c||c|c|c|c|c|c|} \hline
$\Pt^{clust}_{\,cut}$ &\multicolumn{6}{c|}{$\qquad \qquad\Pt^{\,out}_{\,cut}~(GeV/c)\qquad \qquad$}
\\\cline{2-7}
\Gvc&\hbd 5\hbd &\hbd 10\hbd&\hbd 15\hbd&\hbd 20\hbd&\hbd 30\hbd&
\hbd 1000 \hbd \\\hline\hline
    5&    0.000&   -0.002&   -0.001&   -0.002&   -0.002&   -0.002\\\hline
   10&    0.001&    0.000&   -0.002&   -0.002&   -0.002&   -0.002\\\hline
   15&    0.001&   -0.001&   -0.003&   -0.003&   -0.002&   -0.003\\\hline
   20&    0.001&   -0.001&   -0.004&   -0.003&   -0.002&   -0.003\\\hline
   30&    0.001&   -0.002&   -0.004&   -0.004&   -0.004&   -0.004\\\hline
\end{tabular}
\vskip0.1cm
\caption{$\sigma(F),~F= \Fptgj$ }
\label{tab:bi105}
\begin{tabular}{|c||c|c|c|c|c|c|} \hline
$\Pt^{clust}_{\,cut}$ &\multicolumn{6}{c|}{$\qquad \qquad\Pt^{\,out}_{\,cut}~(GeV/c)\qquad \qquad$}
\\\cline{2-7}
\Gvc&\hbd 5\hbd &\hbd 10\hbd&\hbd 15\hbd&\hbd 20\hbd&\hbd 30\hbd&
\hbd 1000 \hbd \\\hline\hline
    5&    0.025&    0.033&    0.035&    0.035&    0.035&    0.035\\\hline
   10&    0.026&    0.036&    0.042&    0.044&    0.046&    0.046\\\hline
   15&    0.027&    0.037&    0.045&    0.052&    0.056&    0.057\\\hline
   20&    0.028&    0.038&    0.046&    0.056&    0.064&    0.067\\\hline
   30&    0.028&    0.038&    0.047&    0.059&    0.076&    0.086\\\hline
\end{tabular}
\end{center}
\end{table}
\def\baselinestretch{0.95}
\begin{table}[htbp]
\begin{center}
\large{ ~$~~\pth= 200~GeV/c$,  ~~~~$\epsilon^{jet}<2\%$.}
\normalsize
\caption{\normalsize Number of events per $L_{int}=3~fb^{-1}$}
\label{tab:bi201}
\vskip-0.0cm
\begin{tabular}{|c||c|c|c|c|c|c|} \hline 
$\Pt^{clust}_{\,cut}$ &\multicolumn{6}{c|}{$\Pt^{\,out}_{\,cut}~(GeV/c)$}
\\\cline{2-7}
\Gvc  &\hcn 5\hcn&\hcn 10\hcn&\hcn 15&\hcn 20\hcn&\hcn 30\hcn&\hcn 1000\hcn \\\hline\hline
    5&      540&     1000&     1090&     1120&     1120&     1130\\\hline
   10&     1380&     3260&     4130&     4480&     4550&     4560\\\hline
   15&     1710&     4200&     5980&     6900&     7410&     7460\\\hline
   20&     1800&     4570&     6730&     8150&     9340&     9650\\\hline
   30&     1840&     4770&     7200&     8970&    11240&    12680\\\hline
\end{tabular}
\vskip0.1cm
\caption{$S/B$}
\label{tab:bi202}
\begin{tabular}{|c||c|c|c|c|c|c|} \hline
$\Pt^{clust}_{\,cut}$ &\multicolumn{6}{c|}{$\Pt^{\,out}_{\,cut}~(GeV/c)$}
\\\cline{2-7}
\Gvc  & 5& 10& 15& 20& 30& 1000  \\\hline \hline
    5&104 $\pm$ 104&177 $\pm$ 167& 114 $\pm$ 84& 116 $\pm$ 85& 116 $\pm$ 85& 116 $\pm$ 85\\\hline
   10&  45.6$\pm$20.0&  51.6$\pm$14.9&  45.4$\pm$11.2&  41.7$\pm$ 9.7&  41.7$\pm$ 9.7&  42.4$\pm$ 9.7\\\hline
   15&  39.8$\pm$14.8&  45.6$\pm$11.5&  41.5$\pm$ 8.4&  33.8$\pm$ 5.8&  30.4$\pm$ 4.9&  30.0$\pm$ 4.8\\\hline
   20&  34.8$\pm$11.9&  40.0$\pm$ 9.1&  33.1$\pm$ 5.7&  27.9$\pm$ 4.1&  23.2$\pm$ 3.0&  21.4$\pm$ 2.6\\\hline
   30&  35.7$\pm$12.2&  33.2$\pm$ 6.9&  27.3$\pm$ 4.2&  22.9$\pm$ 3.0&  19.1$\pm$ 2.1&  16.4$\pm$ 1.6\\\hline
\end{tabular}
\vskip0.1cm
\caption{$\la \Dbgj \ra ~~(GeV/c)$}
\label{tab:bi203}
\begin{tabular}{|c||c|c|c|c|c|c|} \hline\
$\Pt^{clust}_{\,cut}$ &\multicolumn{6}{c|}{$\qquad \qquad\Pt^{\,out}_{\,cut}~(GeV/c)\qquad \qquad$}
\\\cline{2-7}
\Gvc  &\hcd 5\hcd&\hcd 10\hcd&\hcd 15\hcd&\hcd 20\hcd&\hcd 30\hcd&
\hcd 1000 \hcd \\\hline \hline
    5&      4.0&      4.8&      5.3&      5.4&      5.5&      5.8\\\hline
   10&      4.5&      5.9&      6.9&      7.6&      7.8&      7.9\\\hline
   15&      4.6&      6.1&      7.7&      8.9&      9.8&     10.0\\\hline
   20&      4.7&      6.3&      8.0&      9.5&     11.2&     12.0\\\hline
   30&      4.7&      6.3&      8.2&      9.9&     12.8&     15.7\\\hline
\end{tabular}
\vskip0.1cm
\caption{$\la F\ra, ~F=\Fptgj$}
\label{tab:bi204}
\begin{tabular}{|c||c|c|c|c|c|c|} \hline
$\Pt^{clust}_{\,cut}$ &\multicolumn{6}{c|}{$\qquad \qquad\Pt^{\,out}_{\,cut}~(GeV/c)\qquad \qquad$}
\\\cline{2-7}
\Gvc  &\hcd 5\hcd&\hcd 10\hcd&\hcd 15\hcd&\hcd 20\hcd&\hcd 30\hcd&
\hcd 1000 \hcd \\\hline \hline
    5&    0.003&    0.002&    0.003&    0.003&    0.003&    0.003\\\hline
   10&    0.001&    0.002&    0.003&    0.003&    0.003&    0.003\\\hline
   15&    0.001&    0.002&    0.004&    0.004&    0.004&    0.004\\\hline
   20&    0.001&    0.002&    0.003&    0.004&    0.003&    0.003\\\hline
   30&    0.001&    0.002&    0.003&    0.004&    0.003&    0.002\\\hline
\end{tabular}
\vskip0.1cm
\caption{$\sigma(F),~F= \Fptgj$ }
\label{tab:bi205}
\begin{tabular}{|c||c|c|c|c|c|c|} \hline
$\Pt^{clust}_{\,cut}$ &\multicolumn{6}{c|}{$\qquad \qquad\Pt^{\,out}_{\,cut}~(GeV/c)\qquad \qquad$}
\\\cline{2-7}
\Gvc  &\hcd 5\hcd&\hcd 10\hcd&\hcd 15\hcd&\hcd 20\hcd&\hcd 30\hcd&
\hcd 1000 \hcd \\\hline 
    5&    0.015&    0.018&    0.020&    0.021&    0.022&    0.023\\\hline
   10&    0.016&    0.020&    0.023&    0.026&    0.027&    0.027\\\hline
   15&    0.016&    0.021&    0.026&    0.030&    0.033&    0.034\\\hline
   20&    0.016&    0.021&    0.027&    0.031&    0.038&    0.041\\\hline
   30&    0.016&    0.021&    0.027&    0.033&    0.043&    0.050\\\hline
\end{tabular}
\end{center}
\end{table}

\end{document}